\documentclass[prl,aps,groupedaddress,amsmath,twocolumn]{revtex4}

\usepackage{graphicx}
\usepackage{dcolumn}
\usepackage{bm}
\usepackage{color}

\begin{document}




\title{Accelerated ab-initio Molecular Dynamics: probing the weak dispersive forces in dense liquid hydrogen} 
\author{Sandro Sorella}
\author{Guglielmo Mazzola}
\affiliation{International School for Advanced Studies (SISSA) Via Beirut 2,4
  34014 Trieste , Italy and INFM Democritos National Simulation Center,
  Trieste, Italy} 
\email[]{sorella@sissa.it}

 \affiliation{Theoretische Physik, ETH Zurich, 8093 Zurich, Switzerland}
\email[]{gmazzola@phys.ethz.ch}
\date{\today}

\begin{abstract}
We propose an ab-initio molecular dynamics method, capable to 
reduce dramatically the autocorrelation time required for the simulation 
of classical and quantum particles at finite temperature. 
The method is based on an efficient implementation of a first order  
 Langevin dynamics modified by means of a suitable, position dependent 
acceleration matrix $S$.
Here we apply this technique, within a Quantum Monte Carlo (QMC) based wavefunction 
approach and within the Born-Oppheneimer approximation, for determining
the phase diagram of high-pressure Hydrogen with simulations much longer than the autocorrelation time.
With the proposed method, we are able to equilibrate 
in few hundreds steps even close to the liquid-liquid phase transition (LLT).  
 Within our approach  we find that the LLT transition is consistent with recent 
density functionals predicting a much larger transition pressures when the long range 
dispersive forces are taken into account.
\end{abstract}

\maketitle

One of the most important problems in  ab-initio molecular dynamics (MD)  is 
the coexistence of much different time scales underlying physical processes, 
even when computing equilibrium finite temperature properties.
This 
 has been a challenge for  most ab-initio simulations in systems of biophysical interest, the most striking one being protein folding\cite{nolting2005protein,scheraga2007protein},
where the microscopic time scale of molecular vibration is of the order of $fs$, while the macroscopic time scale of the folding exceeds the $\mu s$.
Also in water system simulations, the 
weak Van der Waals (vdW) interactions and the related hydrogen bond imply very difficult equilibration properties at ambient conditions, so that the most accurate simulations are based on force field fitting forces.
We will show here  that, a computationally expensive ab-initio method for dense liquid hydrogen  can be combined with a sampling technique capable to reduce drastically the autocorrelation times. This method has the potential, due to its simplicity, to be applied with success  also to much more complex systems and other ab-initio techniques.
 
At variance of all previous attempts\cite{Bennett_1975,mauri_1994,bussi2007canonical,ceriotti2009langevin,ceriotti2010efficient,mass_2011,john2015quantum,bernemulti}, we propose that an 
optimal way to get rid of different time scales is based on the use of first order Langevin dynamics:
\begin{eqnarray} \label{basic}
\dot { \vec R} &=& S^{-1}(\vec R)  \vec f_{\vec R} (t)  + \vec \eta \nonumber \\
\langle \eta_i \eta_j \rangle&=&2T \delta(t-t^\prime) S_{ij}^{-1}(\vec R)
\end{eqnarray}  
where $T$ is the temperature, $f_{R_i}=-\partial_{R_i} V(\vec R) $ and 
$V_{\vec R}$ is an energy potential of the classical coordinates $\vec R$, e.g.  atomic positions. 
For $S_{ij}=\delta_{ij}$ this is the conventional first order Langevin dynamics  (CFOLD).
Preconditioned Langevin equations, formally similar to Eq.~\ref{basic}, have been already introduced in several  fields different from molecular dynamics, such as gauge field theories\cite{Parisi_1984},  statistics\cite{girolami2011riemann} and machine learning\cite{welling2011bayesian}, with different definitions of the preconditioning matrix $S$. 
Instead, the main idea of this work is based on the simple solution of CFOLD
 for the harmonic potential $V(R) = -K \vec R$, where
$K$ is the elastic matrix term. In this case one is able to show that the 
i) the autocorrelation time is independent of temperature $\tau_{corr}^{-1}=K_{min}$ , where $K_{min}$($K_{max}$) is the lowest (largest) non-zero eigenvalue of the elastic matrix  ii)  for $T=0$ the corresponding discretized equation:
\begin{equation}
\vec R_{n+1}= \vec R_n + \Delta \vec f_{R_n}
\end{equation} 
is nothing but the steepest descent method for the optimization of the potential energy surface $V(\vec R)$.
Thus the autocorrelation time depends just on how fast we are able to approach the minimum possible energy in an energy optimization.

At this point we have to consider  that the steepest descent technique is very much limited by 
the very large condition number 
$K_{cond}=K_{max}/K_{min}$ of the matrix $K$. 
Indeed the number of iterations to approach the minimum is of the order of $ 
\tau_{corr}/ \Delta=K_{cond}$, because the maximum time step $\Delta$,  that can be used in this approach for a stable simulation, is of the order of $\simeq 1/K_{max}$.\cite{mauri_1994}

But why we have to be limited by the steepest descent?
It is well known that, in standard optimization techniques, much better methods exist, and in particular the Newton method:
\begin{equation}
\vec R_{n+1} = \vec R_N + \Delta H^{-1} \vec f_{R_n}
\end{equation}
is able to reach the minimum in {\em one step} for the harmonic case, regardless on how large the condition number is.  In this case $S$ (in Eq.\ref{basic}) 
is given by the Hessian matrix $H={1\over 2} \partial_{R_i} \partial_{R_j} V(\vec R)$,  that is usually much heavy to compute. However it is clear that, by a reasonable choice of $S$, not necessarily given by the Hessian matrix, much better performances of the optimization can be achieved.
 
The key idea of this paper is already clear now. We assume that the autocorrelation time does not depend on $T$, as it is implied by the solution of the 
harmonic case in Eq.(\ref{basic}). Thus, in the general non harmonic case, at finite $T$, by using a position dependent matrix $S$ related to the Hessian one, the autocorrelation time-measured in simulation steps- can be
 drastically reduced 
from $K_{cond}$ to $\simeq 1$\footnote{Notice that, by using second order Newton dynamics rather than first order provides a speed up of $\sqrt{K_{cond}}$, as discussed in Ref.\onlinecite{
mauri_1994}, whereas with the acceleration matrix $S=H$ it is possible to achieve  in principle a much larger speed-up $\simeq K_{cond}$, even compared with standard second order MD.}
 in the 
corresponding discretized version of the LD.

\begin{figure}[h]
\begin{center}
\includegraphics[width=0.67\columnwidth ,angle=-90]{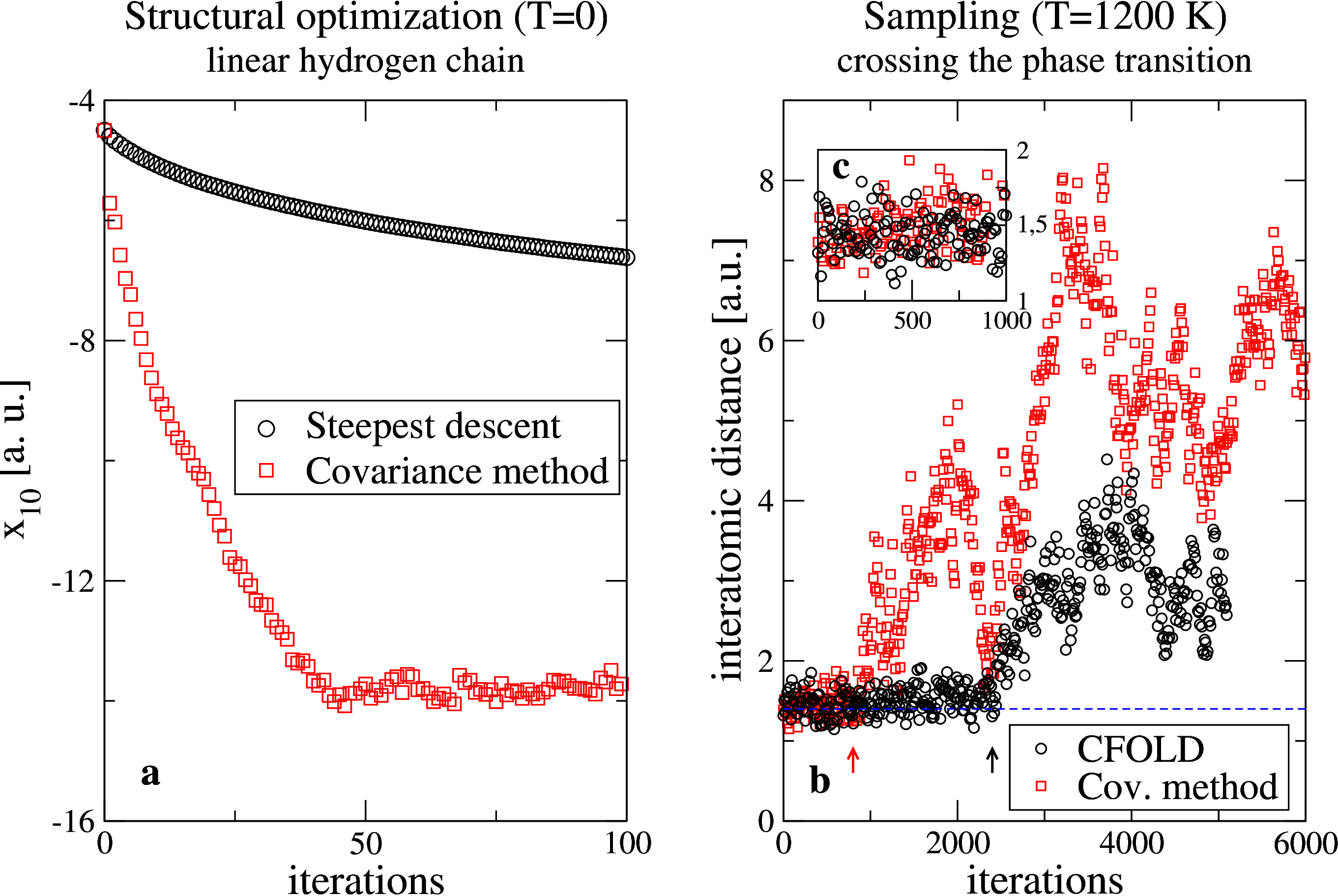}
\caption{ 
\emph{a.)} Position of the left most Hydrogen in the structural optimization of 
a 10 atom Hydrogen system symmetrically confined in a one dimensional open chain.  The starting configuration is obtained by equally spaced Hydrogen at distance $1 a.u.$. The approach to the stable minimum energy configuration  of the steepest descent method and the present covariance method is shown as a function of the number of iterations, at fixed optimal time step $\Delta$ for both cases. $K_{cond} \simeq 200$ in this case.
\emph{b.)} Evolution of the distance between two selected atoms  as the Langevin simulation progresses. At the beginning the two atoms form a molecule, which breaks after several iterations. In red we plot the simulation obtained with the new method, whereas in black with the CFOLD (the starting configuration being equal). The new method displays an enhanced relative diffusion, after the molecule's break (red and black arrows). Nevertheless it integrates with the same accuracy the fast intramolecular motion, at the beginning of the simulation (see inset \emph{c.)}). $K_{cond} \simeq 7$ in this case.
}
\label{fig:cov}
\end{center}
\end{figure}

We adopt a quantum Monte Carlo approach\cite{attaccalite_stable_2008,sorella_algorithmic_2010,mazzola_unexpectedly_2014,zen2015ab},  to perform ab-initio 
simulation of atoms at finite temperature $T$, considered as classical 
particles interacting via the Born-Oppheneimer energy surface, obtained by a quantum mechanical variational optimization of a correlated 
electronic wavefunction containing several parameters:
\begin{equation}
V(\vec R) = {\rm Min}_{\vec \alpha} { \langle \Psi_{\vec \alpha} |H_{\vec R} | \Psi_{\vec \alpha} \rangle \over  \langle \Psi_{\vec \alpha} | \Psi_{\vec \alpha} \rangle } 
\end{equation}
where the variational wavefunction is of the Jastrow-Slater type, 
$\vec \alpha$ indicates generically all the variational parameters,  and $H_{\vec R}$ is the full-many body electronic Hamiltonian with Coulomb interaction, at fixed atomic positions. Since our method is not restricted to QMC we will not enter in  the details of the wavefunction and the optimization methods, the interested reader can refer to our previous works.\cite{casula_geminal_2003,marchi_resonating_2009,zen_molecular_2013,sorella2015geminal}  
In QMC one of the most important matrix that we have found useful for the dynamics is the so called ''covariance matrix'':
\begin{equation}
 Cov(\vec f) = \langle \langle f_i (\vec R) f_j (\vec R) \rangle \rangle - \langle \langle f_i (\vec R) \rangle \rangle \langle \langle f_j (\vec R) \rangle \rangle 
\end{equation}
where $\langle \langle~~ \rangle \rangle$ indicates a statistical average over a given number of samples at fixed atomic positions. 
This covariance matrix (at variance of the Hessian one) is always positive 
definite, and empirically it has been shown,  in several system cases,
 to be almost proportional to the Hessian matrix\cite{yeluo_2014}, at least at the equilibrium structure, where the Hessian is also positive definite.
For this reason it is natural to take the covariance matrix 
for accelerating the LD within QMC, and we have assumed $S=  Cov(\vec f)$ in the following.
Indeed it is possible to show that 
the direction 
$\vec s = Cov(\vec f)^{-1} \vec f$
represents just the one with maximum 
signal to noise ratio for the corresponding force energy derivative $ \vec s \cdot \partial V / d \vec R$. 
This explains why it is more likely to find a lower energy in this direction, 
clearly suggesting its importance for QMC energy minimization.
With this choice the Eq.(\ref{basic}) is also covariant, namely independent of an arbitrary
change of coordinates $\vec R \to \vec R^\prime (\vec R)$.
That this represents a good choice is shown in a simple Hydrogen system 
in Fig.~\ref{fig:cov},
 where $K_{max}/K_{min}$ is rather large due to a weak 
molecular binding as opposed to the large molecular 
frequency of  $H_2$.
 Within the present method the equilibrium positions of 
this difficult molecule
 are reached within a few dozens iterations, whereas it is not possible to approach the stable configuration with the standard steepest descent method.

\begin{figure}[h]
\begin{center}
\includegraphics[width=0.5\columnwidth ,angle=-90]{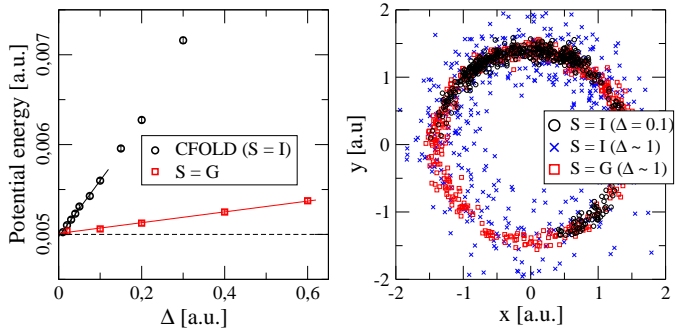}
\caption{ 
\emph{Left panel:} Average potential energy as a function of the integration time-step $\Delta$ for a simple toy model (see SI for details). Black points refer to the standard first order Langevin dynamics (setting $S = I)$), while red points to the improved one, with non-trivial $S$, given by the metric $G$ 
of the  corresponding non linear space (see SI).
\emph{Right panel:} Configuration sampled with a fixed number of iterations (500) by the two different dynamics.
Using the new dynamics (red square points) we can employ a large time step ($\Delta \approx 1$) to accelerate the sampling of the slow rotational degree of freedom while maintaing the same accuracy of a CFOLD having a ten times smaller $\Delta$ (black circles). The CFOLD with $\Delta \approx 1$ is instead unstable (blue stars).
}
\label{fig:model}
\end{center}
\end{figure}

At finite temperature the discretization of the Langevin equation for a
finite time step $\Delta$  is highly non trivial, and defines a Markov chain with unique equilibrium distribution.
By requiring that, for $\Delta \to 0$, the equilibrium distribution is just the 
canonical one $f(\vec R) \propto \exp{-{V(\vec R) / T}}$ it is possible 
to show that (see supplementary informations SI) the following iterative 
scheme:
\begin{eqnarray} \label{finaldyn}
\vec R (t+\Delta) &=& \vec R(t) +\sqrt{2 T \Delta} \vec z (t) +S^{-1}(\vec R) \left\{ \Delta \vec f_{\vec R} \right. \nonumber \\
&-& \left. \left[ { S(\vec R(t-\Delta))- S(\vec R) \over 2}\right]  (\vec R(t-\Delta) -\vec R(t)) \right\} \nonumber \\
\langle z_i (t) z_j (t) \rangle &=& S^{-1}_{i,j} (\vec R(t)) 
\end{eqnarray}
fulfills the equilibrium target.

  We remark here  that
 the main advantage of  Eq.(\ref{finaldyn}) is that it takes  
 into account the explicit dependence of the matrix $S$ on the atomic positions $\vec R$, without using any cumbersome derivative of its inverse,  
as unavoidable in other methods\cite{risken,mass_2011,mazzola_finite-temperature_2012}, that are computationally  much more expensive.
This iteration scheme
 is not a Markov chain as the positions $\vec R$ at the 
next time $t+\Delta$  depend not only on the actual time $t$ but also on the previous ones $t-\Delta$. However,  in the limit of $\Delta \to 0$ it can be shown that this convenient iteration scheme is 
 equivalent to the much more 
complicated Markov chain used in our previous work\cite{mazzola_finite-temperature_2012}. 



In order to warm up with the capabilities of the method we show in Fig.~\ref{fig:model} its performances in a toy model where only two time scales are present (vibrational and rotational). The clear advantages of the method are evident also when the 
acceleration matrix is not exactly equal to the Hessian one (see SI for details).
In this case we can accurately integrate the fast degrees of freedom while speed-up the sampling along the slowly varying direction. This property holds also in the more realistic case of liquid hydrogen, where we can accurately sample the intramolecular vibrations while accelerating the inter-molecular diffusion. (see Fig.\ref{fig:cov}).


\noindent {\bf Liquid-liquid transition in dense hydrogen.}
 Extensive experimental as well as theoretical efforts have been devoted to understand the high pressure phase diagram of hydrogen, the simplest possible condensed matter system in nature.
One among many open problems regarding the behavior of this compound at high pressures is finding  atomization in the liquid sector of the phase diagram, i.e. the boundary between the molecular and the atomic liquid at higher pressures, which must have also metallic character. 

Until few years ago, all the few experimental observations\cite{weir_metallization_1996,fortov_phase_2007,dzyabura_evidence_2013} located the insulator to metal transition (IMT), which provides a lower bound for the molecular to atomic transition, at pressures of $\sim$ 140 GPa in the temperature range of 1500-3000 K.
Concurring numerical simulations\cite{scandolo_liquidliquid_2003,tamblyn_structure_2010,morales_evidence_2010}, using both DFT, with Perdew-Burke-Ernzerhof (PBE) exchange-correlation functional\cite{perdew_generalized_1996}, and QMC agreed with this value.
In Ref.~\onlinecite{mazzola_unexpectedly_2014} we observed instead the atomization at $\sim$ 350 GPa and 2300 K and around 600 GPa at 600 K, with simulations 
of $256$ hydrogen atoms at the $\Gamma$ point.

Very recently, two new experiments have been performed to address conclusively this issue, but they provided two very different pictures.
In the first, done by Silvera and coworkers\cite{zaghoo2016evidence} a first order IMT in hydrogen in a pressure range which agrees quantitatively with previous experiments and simulations. 
This evidence has been qualitatively confirmed also by Ref.~\onlinecite{ohta2015phase}, although this experiment focus on larger temperatures.
In a second experiment, Knudson and coworkers\cite{knudson2015direct} observed the IMT in deuterium at much larger pressures instead. Their IMT phase boundary line is almost vertical in the $P-T$ phase diagram and is located at around 300 GPa in a temperature range between 1200 and 1800 K.
Since the two experiments were performed at temperature as high as 1800 K, it is very unlikely that this huge difference is due to the enhanced zero-point motion of hydrogen compared to deuterium\cite{weir_metallization_1996}.
We notice that this new transition pressure would be now  much more in agreement with DFT simulations using non-local exchange-correlation functionals\cite{morales_nuclear_2013} such as the vdW-DF1 and DF2, belonging to the vdW class of functionals, introduced in Refs.~\onlinecite{vdwdf1,vdwdf2}.


We apply this new  framework to liquid hydrogen, using a carefully prepared simulation set-up. At variance of Refs.~\cite{mazzola_unexpectedly_2014,mazzola2015distinct} we take care of finite size-effects using a recently developed\cite{refkturbo} $k-$ point sampling of the Brillouin zone\cite{pierleoni_coupled_2004,foulkes_kpoint95} (BZ).
In most simulations we use a cubic supercell containing 64 atoms and a 4$\times$4$\times$4 Monkhorst and Pack k-point mesh, which looks clearly consistent 
with  larger systems calculations (see Fig. 8 in SI).
We perform simulations at fixed volumes and temperatures, from 900 to 1800 K.
Following Ref.~\onlinecite{mazzola_unexpectedly_2014} we identify the transition density for each temperature by tracing the discontinuity
in the $P$ vs $\rho$ equation of state as well as the jumps of the radial pair distribution function (see SI).
 In order to check the 
 convergence of the MD  to the canonical equilibrium distribution, 
for each point of the phase diagram we 
start the NVT simulations, using two configurations, generated almost randomly,
 with molecular and atomic characters (see SI).
 Away from the atomic-molecular coexistence region we were able to obtain 
consistent and statistically converged results  with at most a few hundreds iterations.  Remarkably we have noticed that, close to the transition, the autocorrelation time increases substantially and the system can easily
be trapped in a metastable phase for 
short simulations.
Our present method can give a meaningful speed-up in this case, but obviously does not solve the problem to cross easily a free energy barrier between two different phases, and this explains why several thousands iterations are required in the coexistence region. We notice, however, that for this problem several other methods are available, e.g. metadynamics\cite{laio2002escaping}, and they can be obviously combined with the present scheme.

Moreover we have found a meaningful dependence of the phase diagram 
on the number of optimization steps $n_{opt}$ used to 
approach the BO energy surface. Indeed we have observed that, even though the 
total energy converges very quickly with $n_{opt}\simeq 7$ (the choice in Refs.~\onlinecite{mazzola_unexpectedly_2014})
 to the lowest variational energy, the ionic forces may require much 
 more optimization steps and therefore the equilibrium distribution can significantly  change even without an apparent improvement in the energy.  
In any event the convergence is exponential in $n_{opt}$ and we have not observed any meaningful change after $n_{opt}\ge 15$, thus reporting here 
 the converged phase diagram for $n_{opt}=20$.
\begin{figure}[h]
\begin{center}
\includegraphics[width=0.7\columnwidth ,angle=-90]{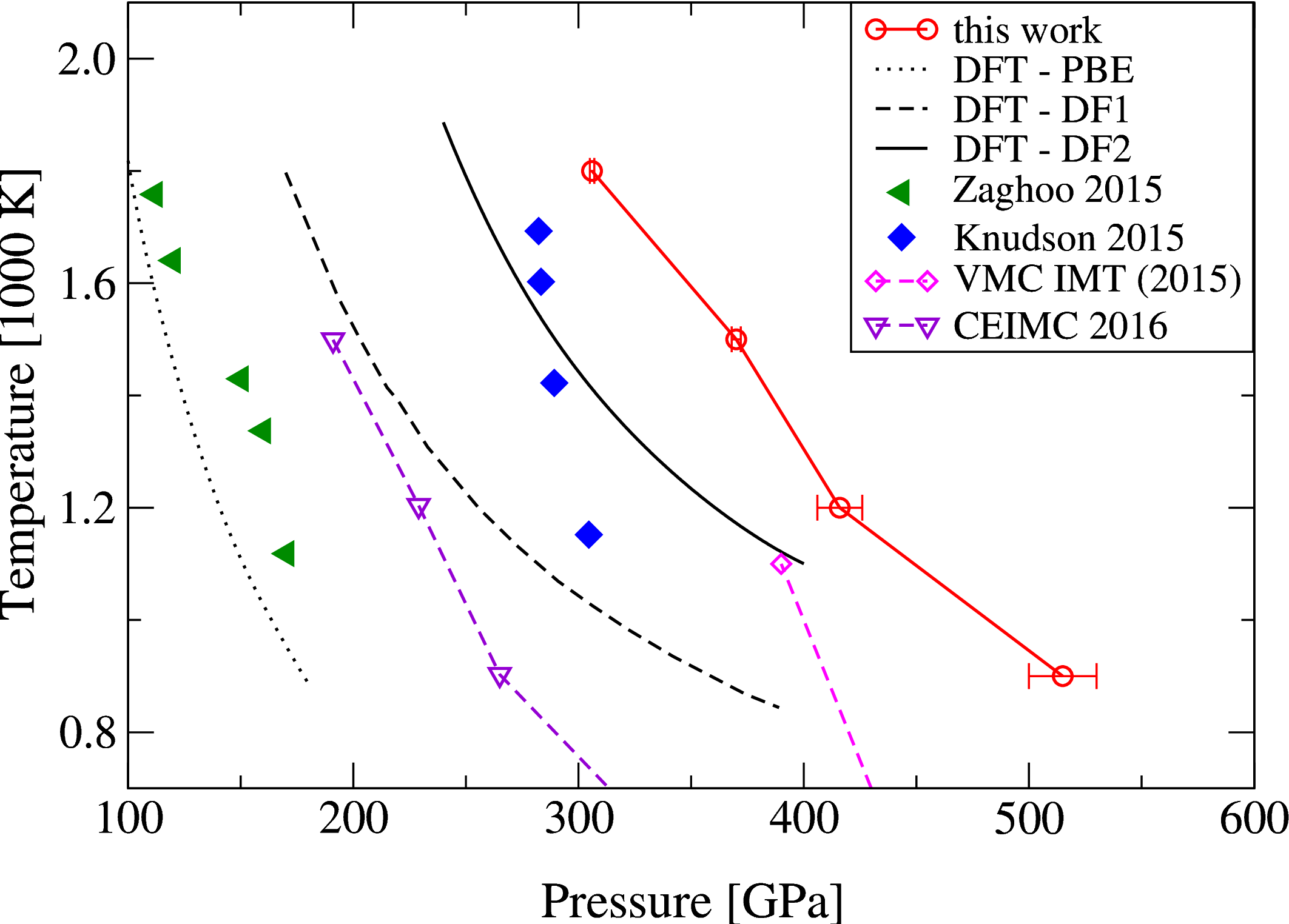}
\caption{ 
Phase diagram of liquid hydrogen. We indicate experimental results of Refs.~\onlinecite{zaghoo2016evidence,knudson2015direct} with solid symbols, while theoretical results (with classical nuclei approximation) of Refs.~\onlinecite{ pierleoni2016liquid, mazzola2015distinct} with open symbols. The present results are plotted in red. We also plot DFT metalization lines using three popular exchange correlation functionals, PBE, vdW-DF1 and DF2. These results are taken from Ref.~\onlinecite{knudson2015direct}.
}
\label{fig:phase}
\end{center}
\end{figure}

The phase boundary does not appear very sensitive to the quality of the trial wavefunction (WF) of the Jastrow-Slater type here employed. In this work we use two kinds of WFs, which have different Jastrow factors. This part takes into account dynamical electronic correlations, and, depending on its explicit form, may represents less or more accurately the  weak long range forces (see SI). 
In particular, in our scheme, the most  accurate description is obtained when 
 the Jastrow operator contains also a  4 body (4B J) electron-electron-ion-ion interaction, whereas  a cheaper but less accurate description is  given 
by using only a smaller (3B J)  factor containing only a three  body electron-electron-ion correlation term.

Our prediction for the  phase transition is now much closer to the recent DFT functionals with 
empirical dispersive forces, vdW-DF1 and DF2, but remains at much larger pressures than 
other  theoretical approaches\cite{morales_evidence_2010,pierleoni2016liquid}, that are based 
on DFT functionals without dispersive forces or on quantum Monte Carlo methods without  MD.

We notice that these results, at low temperatures,  are compatible with our metallization prediction of Ref.~\onlinecite{mazzola2015distinct}, though  the  observed residual molecular fraction in the metallic phase, may be an artifact of the $\Gamma$ point BZ-sampling.
Moreover, the effective inclusion of weak dispersion forces, by means of the 4B Jastrow term, does not significantly change the transition pressure, though providing 
a lower internal energy  of about  $1mH/atom$ both  in the molecolar and the atomic phases. 
It is not the purpose of the present paper to resolve the present discrepancies between different approaches, but we nicely observe that the theoretical uncertainty is becoming smaller,  in the  very recent works (including this one).

In conclusions we have proposed a general  method of accelerating the present ab-initio MD, that is simple enough and can be quite generally adopted in all ab-initio schemes, because the inversion of a  matrix $S$, whose leading dimension is the number of ion coordinates, represents just a negligible  overhead in these cases\footnote{Ab-initio methods scales ad $N^3$, when $N$ is the electron number, and the solution of a linear problem is one of the fastest and efficient $N^3-$ scaling algorithm, known in computer science.}, and 
the computational gain, for suitable acceleration matrices, can be in principle of several orders of magnitude, proportional to $K_{cond}$.
In our QMC based approach we have used an acceleration matrix related to the correlation
of the noise in the VMC force components, namely the covariance matrix. In general we expect 
that considerable gains in efficiency can be obtained by using for $S$ 
 the Hessian matrix $H$ determined by accurate and cheap empirical potentials, at most 
corrected as $S=H+\mu I $ with $\mu>0$ suitable chosen when $H$ is not positive definite 
for some atomic positions $\vec R$.   
In this work we have not exploited this possibility that could be in principle much better 
than our present choice,  
because $S$ does not need to be estimated stochastically (see SI for a simple model).
This method can be easily extended\footnote{In preparation.} to the quantum case by using the mapping of a finite temperature quantum simulation to an extended classical system\cite{RevModPhys.67.279,marx_1996}.

We have applied this method to a topic of recent interest, by providing well 
converged results on the liquid-liquid phase transition, within the Jastrow-Slater variational ansatz,  and found good agreement  with a recent experiment (at $T=1800K$, see Fig.\ref{fig:phase})\cite{knudson2015direct}  and DFT results  empirical dispersive forces functionals.
The classical nuclei approximation here adopted may explain the residual difference compare to this experiment. 
In particular, we expect  the slope of the transition line in the $P-T$ phase diagram to become more vertical including  quantum nuclear effects, which become more important as the temperature decreases.
This is a first simple application, where the problem of different time scales is not even particularly important ($K_{cond}\simeq 7$)  and we expect much more dramatic speed up in polyatomic systems such as 
water, where the slow dynamics of the Hydrogen bond network contrasts by several orders of magnitude the high frequencies vibrations of the water monomer 
($K_{cond} \simeq 2000$). 
As this manuscript was prepared, an experimental study was published in Ref.~\onlinecite{PhysRevLett.116.255501}, ruling out the possible first order phase transition at low pressures, and clearly supporting 
 our simulations, as well as DFT studies with non-local vdW functionals and the earlier experiments in 
Ref.~\onlinecite{knudson2015direct},
all  predicting much higher liquid-liquid dissociation transition pressures in the range 250-300 GPa at $\sim$ 2000 K. 

\emph{Acknowledgments.}
We acknowledge very useful comments to the manuscript 
by M. Ceriotti and M. Parrinello.
We also acknowledge A. Zen for the calculation of $K_{cond}$ 
 in the water dimer.
G.M. was supported by  the European Research Council through ERC Advanced Grant SIMCOFE, and the Swiss National Science Foundation through the National Competence Centers in Research MARVEL and QSIT. Computational resources were provided by  AICS-Riken and by CINECA in Bologna.

\clearpage
\onecolumngrid
\appendix

\section{ Proof of  equilibration to the canonical distribution }
In the first section of this Supporting Information we provide the formal 
derivation of our iteration scheme.
By means of an highly non trivial 
time discretization of  the proposed  Langevin dynamics it is possible 
 to sample the canonical distribution $P(\vec R,t) ={ \exp(- V( \vec R ) /T ) \over Z}$ 
 in an efficient way.
As well known the integration of the Langevin equation in Eq.(1) of the manuscript is not univocally defined 
 when the matrix $S$ is explicitly dependent on $\vec R$.
In the following we use a pragmatic point of view and define a discretization 
of the Langevin equation in a short time interval $\Delta$, by requiring 
that for $\Delta \to 0$ the associated Markov chain equilibrates to  the canonical distribution.

The usual discretization of Eq.(1), namely 
evaluated at discrete time $t=n \Delta$,
can be found in textbooks\cite{risken}  and has the rather involved expression 
\begin{eqnarray} \label{standard}
\vec R (t+\Delta)_j &=&  \vec R (t)_j + \Delta \left(S^{-1}(\vec R) \vec f_{\vec R}  \right)_j 
 +T \Delta \left( \sum\limits_{i} \partial_i S^{-1}_{ji}(\vec R)   \right) 
 +\sqrt{2 T \Delta} z_j(t) \nonumber \\
\langle z_i (t) z_j (t) \rangle &=& S^{-1}_{i,j} (\vec R(t)) \\
\left( \vec f_{\vec R}\right)_j&=& -\partial_j V(\vec R)  +\gamma T \partial_j \ln |S(\vec R)|
\end{eqnarray}
where $N$ is the dimension of the real vector $\vec R$ and 
 $\gamma=1/2$ if the covariant metric is used in the definition of the 
partition function:
\begin{equation}
 Z = \int dR^N |S(\vec R)|^\gamma  \exp \left(- {V(\vec R)\over T}  \right)
\end{equation}  
where $|S|$ stands for the determinant of the symmetric and real matrix $S$.
In the following we are interested to the canonical distribution, defined in the standard Euclidean metric, and therefore we have to keep in mind 
in the following that $\gamma=0$, the forthcoming derivation being valid 
for any value of $\gamma$.

The above Markov chain defines a discretized Master equation for the 
probability function $P(\vec R,t)$, that in the limit $\Delta \to 0$ 
becomes a Fokker-Planck equation of the following form:

\begin{equation} \label{fokker}
\partial_t P(\vec R,t)= \sum\limits_j \partial_j 
\left \{ - \left[ S^{-1}(\vec R) f_{\vec R} \right]_j  P(\vec R,t)  +T 
\sum\limits_i \left[ 
S^{-1}_{j,i}(\vec R) \partial_i   \right]    
P(x,t) \right\}
\end{equation}
In order to find the equilibrium distribution it is enough to equate to zero 
the term between braces which immediately gives $P(\vec R,t)$ up to a constant,
that is in turn is determined by the normalization condition 
of probabilities, yielding:
\begin{equation}
P_{eq} (\vec R,t) ={ |S(\vec R)|^\gamma \exp (- V( \vec R) /T ) \over Z },
\end{equation} 
namely the desired distribution.
So far so good. Unfortunately the Markov chain of Eq.(\ref{standard}) is 
not practical, because it contains the ''cumbersome term'':
\begin{equation}
\label{e:cumber}
{\Gamma}_j(\vec R) =\sum\limits_i \partial_i  S^{-1}_{j,i}(\vec R)
\end{equation}
Indeed the calculation of the inverse of a matrix takes the order of $N^3$ 
operations, as well as, clearly each derivatives over any variable $i$ 
(e.g. by the finite difference method). 
In order to make the summation over all $i$ for each $j$ in the above equation
we end up with an algorithm scaling in most cases as the fourth power of $N$, unless for particularly simple cases.
Moreover the expressions for the inverse derivatives become so much complicated that are very difficult to implement in practice, especially within the QMC approach.

\subsection{ First simplified Iteration scheme }
Before deriving the final convenient expression for sampling in the 
most efficient way the canonical distribution by the proposed 
accelerated Langevin dynamics, we consider the following 
Markov chain that does not require the evaluation of the ''cumbersome term'' $\Gamma_j (\vec R)$:
\begin{eqnarray} \label{smartMarkov}
\vec y &=& \vec R+ \sqrt{2 T \Delta} \vec z(t) \nonumber \\ 
\vec R^\prime &=& \vec y + \Delta S^{-1}(\vec R) \vec f_{\vec R} - {1 \over 2} S^{-1}(\vec R) \left[ S(\vec y)- S (\vec R) \right]  (\vec y -\vec R) 
 \nonumber \\
\langle z_i (t) z_j (t) \rangle &=& S^{-1}_{i,j} (\vec R) \label{noisez}
\end{eqnarray}
where here $\vec R=\vec R(t)$ and $\vec R^\prime=\vec R(t+\Delta)$.
We will show in the following that the above Markov chain implies the 
same Fokker-Planck equation (\ref{fokker}) corresponding  
the much more involved 
 discretization in Eq.(\ref{standard}) and therefore the equilibrium distribution for $\Delta\to 0$ will be the correct one. 
For any non zero $\Delta$ we can use a very well established result of Markov chains, that can be seen as an extension of the Perron-Frobenius theorem to non negative arbitrary matrices, and stating that, even in this case a unique equilibrium distribution is reached nearby ($O(\Delta)$) to the desired one, even though the detailed balance condition for the conditional probability, 
i.e. $ K(\vec R^\prime| \vec R) P_{eq}(\vec R) = K(\vec R| \vec R^\prime) P_{eq} (\vec R^\prime)$,  is not satisfied\cite{bookstat}. 
This shows  that it is possible to avoid the ''cumbersome term'', with a minor computational effort, that in this case  amounts to calculate the matrix $S$ two times for each time step.

Let's therefore proceed with  the main proof of this section.
The Markov chain in Eq.(\ref{noisez}) 
defines in a unique way the conditional probability density of having 
$ \vec R^\prime =\vec R(t+\Delta) $ given $\vec R (t)=\vec R$:
\begin{equation}
K( \vec R^\prime | \vec R) = { \int dz^N \exp\left[ -{1 \over 2 } (z , S(\vec R) z) \right] 
\delta\left\{ \vec R^\prime - \vec R -\Delta  S^{-1} (\vec R) \vec f_{\vec R} 
 -\sqrt{2 T \Delta }\vec z +{1 \over 2} S^{-1}(\vec R) \left[ S(\vec y)- S (\vec R) \right]  (\vec y -\vec R) \right\} 
\over \int dz^N \exp\left[ -{1 \over 2 } (z , S(\vec R) z) \right]  } \label{defK} 
\end{equation} 
Here and henceforth we denote by $(a,b)$ the scalar product of two $N-$ dimensional real vectors.
We want to obtain a Fokker-Planck equation in the limit of $\Delta \to 0$.
To this purpose we write the Master equation:
\begin{equation}
P_{n+1}( \vec R^\prime) = \int dR^N K(\vec R^\prime | \vec R) P_n( \vec R) 
\end{equation}
and employ the integration in $dR^N$ after substituting the expression 
of $K(\vec R^\prime|\vec R)$ given above.
For this purpose we solve the argument of the $\delta$ function, by replacing 
$\vec R$ with $\vec R^\prime$ when it is allowed at the leading 
order in $\Delta$:
\begin{equation} \label{transf}
\vec R(\vec R^\prime)= \vec R^\prime -\sqrt{2 T \Delta} \vec z - \Delta S^{-1} (\vec R^\prime) \vec f_{\vec R^\prime} 
+{1 \over 2} S^{-1}(\vec R^\prime) \left[ S(\vec R^\prime+\sqrt{2T\Delta} \vec z)- S (\vec R^\prime) \right]  \sqrt{2T\Delta} \vec z+ o(\Delta)  
\end{equation}
and that:
\begin{equation}
\int dz^N \exp\left[ -{1 \over 2 } (z , S(\vec R) z) \right] = (2 \pi)^{N/2} \exp 
\left[-{1\over 2}  Tr[ \ln S(\vec R) ] \right]
\end{equation}

We obtain therefore that the Master equation for the evolution of the probability is explicitly given:
\begin{equation} \label{iter1}
P_{n+1} (\vec R^\prime ) =\int  dz^N J_{\Delta} (\vec R(\vec R^\prime))  { (2 \pi)^{N/2}  \exp \left[ -1/2 
(z,S\left(\vec R(\vec R^\prime)\right) z) \right] P_n\left(\vec R(R^\prime) \right)
   \over \exp \left[-{1\over 2}  Tr[ \ln S\left(\vec R(\vec R^\prime)  \right) ] \right] } 
\end{equation}
where  $J_{\Delta} (\vec R(\vec R^\prime))=1+\Delta B(\vec R^\prime)+o(\Delta)$ 
is the Jacobian of the transformation of Eq.(\ref{transf}), that can be expanded in $\Delta$ with a  well defined expression for $B(\vec R^\prime)$ that we do not explicitly write in the following, because, as we will see soon, it is not important for the derivation. 

Indeed,  by substituting the transformation of Eq.(\ref{transf}) in 
 Eq.(\ref{iter1}), and expanding the latter equation  to the leading order in $\Delta$  we obtain the following expression:
\begin{eqnarray}
P_{n+1} \left( \vec R^\prime \right)&=& 
\left[1 +\Delta C(\vec R^\prime) \right] P_n (\vec R^\prime)  
+ \int dz^N \mu_{\vec R^\prime} (\vec z) 
\Bigg\{  
  \nonumber \\
&-&\sum\limits_j  
\left\{ \sqrt{2 T\Delta} z_j  + \Delta 
\left[ S^{-1}(\vec R^\prime )  \left( \vec f_{\vec x^\prime}  
-{S(\vec R^\prime+\sqrt{2T\Delta} \vec z)- S (\vec R^\prime) \over 2 \Delta }  \sqrt{2T\Delta} \vec z \right) \right]_j
 \right\}  \partial_j P_n( \vec x^\prime) \nonumber \\
&+&
\left. \Delta T  \sum\limits_{i,j} z_i z_j \partial_i \partial_j P_n(\vec R^\prime)  \right\} + o(\Delta) \label{iterate}
\end{eqnarray}
where $\mu_{\vec R^\prime}  (\vec z)$ 
is the probability density for the random 
vector $\vec z$:
\begin{equation}
\mu_{\vec R^\prime}(\vec z) =  (2 \pi)^{N/2} \exp \left\{ -1/2 
\left[ (z,S\left(\vec R^\prime \right) z) -Tr[ \ln S\left(\vec R^\prime \right) ] \right]  \right\}
\end{equation}
that is normalized because  $\vec R^\prime$ is given and does 
not dependent on $\vec z$.

In the above iteration in Eq.(\ref{iterate}) there is therefore a term 
that simply multiplies $P_n(\vec R^\prime)$ by a function:
\begin{equation}
1+ \Delta C(\vec R^\prime)
\label{firstterm}
\end{equation}
where $C(\vec R^\prime)$ is rather involved and comes from the expansion 
in small $\Delta$ of all the integrand in Eq.(\ref{iter1}):
\begin{equation} \label{defC}
C(\vec R^\prime)\Delta= 
-1+\int dz^N  J_\Delta ( \vec R(\vec R^\prime)) \mu_{\vec R (\vec R^\prime)} (\vec z)= \Delta B(\vec R) -1 + \int dz^N \mu_{\vec R (\vec R^\prime)} (\vec z)+
o(\Delta)
\end{equation}
Notice that $\vec R(\vec R^\prime)$ depends on the 
random variable $\vec z$ via Eq.(\ref{transf}) and therefore the term 
$\int dz^N \mu_{\vec R (\vec R^\prime)} (\vec z)$ 
in the above equation is non trivial and different from $1$ by $O(\Delta)$. 
We will not attempt to calculate this term, as well as $B(\vec R)$, 
but derive it from the 
conservation of the normalization condition of the probability.

The term that couples to the first derivative of $P_n( \vec R^\prime)$
reads:
\begin{eqnarray} 
 &-&\Delta \sum\limits_j \left[ S^{-1} 
\left( \vec f_{\vec R^\prime}-{S(\vec R^\prime+\sqrt{2T\Delta} \vec z)- S (\vec R^\prime) \over 2 \Delta } \sqrt{2T\Delta} \vec z \right) \right]_j \partial_j P_n( \vec x^\prime ) \label{casino} \\
& -&\int dz^N \mu_{\vec R^\prime}  (\vec z) \sum\limits_{i,j,k,l} (2 \Delta T) z_j \left[ {1\over 2} 
z_k \left(\partial_i S_{k,l} (\vec R^\prime) \right) z_l z_i  -{1\over 2 } S^{-1}_{k,l} (\vec R^\prime)  \left(\partial_i S_{kl} (\vec R^\prime) \right) z_i \right]  \partial_j P_n( \vec x^\prime )  \label{secondall}
\end{eqnarray}

In the above equations we have also used the following relation:
\begin{eqnarray}
Tr[ \ln S\left(\vec R(\vec R^\prime)\right] &=& Tr[ \ln S(\vec R^\prime)] 
-\sum\limits_i 
Tr [S^{-1}(\vec R^\prime) \partial_i S(\vec R^\prime) ]  \sqrt{2 T\Delta} z_i +O(\Delta)  \nonumber \\
&=& Tr[ \ln S(\vec R^\prime)] -\sqrt{2 T\Delta} 
\sum\limits_{i,k,l} S^{-1}_{kl} (\vec R^\prime) \partial_i S_{kl} (\vec R^\prime)  z_i + O(\Delta) 
\end{eqnarray}
coming from the expansion of $\mu_{\vec R(\vec R^\prime)} (\vec z)$, and 
in the last equation we have used that $S^{-1}$ is symmetric because $S$ is symmetric.
By carrying out the simple integration in $dz^N$, i.e. by replacing 
$\int dz^N \mu_{\vec R^\prime} (\vec z)  z_j=< z_j > =0$ and 
$\int dz^N \mu_{\vec R^\prime} (\vec z)  z_i z_j= < z_i z_j> 
=S^{-1}_{i,j} (\vec R^\prime)$ 
and by applying the Wick's theorem for the integration of the higher order 
polynomial involved, i.e.  $ < z_j z_k z_l z_i > = < z_j z_k > < z_l z_i > + 
< z_j z_l> < z_k z_i> + <z_j z_i> <z_k z_l>$, 
 we obtain that  
Eq.(\ref{secondall}) reads:
\begin{eqnarray} \label{secondpartial}
-\Delta T \sum\limits_{i,j,k,l} 
\left[ \left( S^{-1}_{jk} S^{-1}_{li} 
+S^{-1}_{jl} S^{-1}_{ki} + S^{-1}_{ji} S^{-1}_{kl} - S^{-1}_{kl} S^{-1}_{ji} \right)  \partial_i S_{kl} \right] (\vec R^\prime) \partial_j P_n( \vec R^\prime)
& =& 2 \Delta T \sum\limits_{i,j} \left(\partial_i S^{-1}_{j,i}(\vec R^\prime)\right) \partial_j P_n(\vec R^\prime) \nonumber \\
&=& 2 T \Delta \sum_j \Gamma_j (\vec R) \partial_j P(\vec R^\prime)
\end{eqnarray}
Where, by $ S+\delta S= S(I+S^{-1} \delta S) \to (S+\delta S)^{-1}= 
S^{-1}- S^{-1} \delta S S^{-1} +o(\delta S) \to \partial_i S_{ji}^{-1}= -\left[ S^{-1} (\partial_i S ) S^{-1} \right]_{ji}$, 
 we easily verify that the LHS and RHS 
of the above Eq.(\ref{secondpartial}) are consistent, as, for instance,
 by using that $S$ is a symmetric 
matrix, we have that: 
\begin{equation} \label{defsijder}
\sum_{kl} S^{-1}_{jl}(\vec R^\prime)  S^{-1}_{ki}(\vec R^\prime)  \partial_i S_{kl}(\vec R^\prime)  =  
\sum_{kl} S^{-1}_{jl}(\vec R^\prime)  S^{-1}_{ki} (\vec R^\prime) \partial_i S_{lk}(\vec R^\prime)  =  
-\partial_i S^{-1}_{ji}(\vec R^\prime)  
\end{equation}
Thus this term partially cancels with the contribution 
 coming from the 
expansion in small $\Delta$ of the term 
\begin{equation}
\left[S(\vec R^\prime+\sqrt{2T\Delta} \vec z)- S (\vec R^\prime) 
\right]_{k,l}
= \sqrt{2 T\Delta } \sum\limits_i  \partial_i S_{kl} (\vec R^\prime)  z_i  
\end{equation}
Indeed, in the Fokker-Planck equation, the term proportional to $\partial_j P$
coming from the Eq.(\ref{casino}) acquires a contribution:
\begin{equation} \label{questa}
T \Delta 
\int dz^N \mu(\vec z) \sum_{i,k,l} S^{-1}_{j,k} (\vec R^\prime) 
\partial_i S_{k,l} (\vec R^\prime) z_i z_l=
T \Delta \sum_{i,k,l}  S^{-1}_{j,k}(\vec R^\prime)  \partial_i S_{k,l} (\vec R^\prime) S^{-1}_{i,l}(\vec R^\prime) =-T\Delta 
\sum_i \partial_i S^{-1}_{j,i}(\vec R^\prime)  = -T \Delta \Gamma_j (\vec R^\prime)
\end{equation} 
where in the last equality we have used the relation given in Eq.(\ref{defsijder}). 
Thus the total term proportional to $\partial_j P_n (\vec R^\prime)$ reads:
\begin{equation}\label{secondtotal}
 \sum\limits_j \left\{ -\Delta \left[ S^{-1} (\vec R^\prime) 
 \vec f_{\vec x^\prime}\right]_j 
 +T \Delta  \Gamma_j (\vec R^\prime) \right\}  \partial_j P_n(\vec R^\prime) 
\end{equation}

Finally the term proportional to the second derivative leads to:
\begin{eqnarray} \label{thirdterm}
\Delta T \sum\limits_{i,j} S^{-1}_{ij}(\vec R^{\prime}) \partial_i \partial_j P_n(\vec R^\prime) &=& T \Delta \sum\limits_i \partial_i \left[ S^{-1}_{ij}(\vec R^{\prime})  \partial_j  P_n( \vec R^\prime)\right] - T \sum\limits_j \left[ \sum\limits_i \partial_i S^{-1}_{i,j} (\vec R^\prime) \right] \partial_j P_n(\vec R^\prime) \nonumber \\ 
  &=& T \Delta \sum\limits_i \partial_i \left[ S^{-1}_{ij}(\vec R^{\prime})  \partial_j  P_n( \vec R^\prime)\right]  -T \Delta \sum_j \Gamma_j (\vec R^\prime) \partial_j P_n(\vec R^\prime)
\end{eqnarray}
By collecting all the terms obtained in Eqs.(\ref{firstterm},\ref{secondtotal},\ref{thirdterm}) all the terms proportional to the ''cumbersome one'' $\Gamma_j( \vec R^\prime)$ cancel out and, by  carrying out the limit $\Delta \to 0$ we obtain the following Fokker-Planck equation:
\begin{equation}
\partial_t P(\vec R,t)= \sum\limits_j \partial_j 
\left \{ - \left[ S^{-1} f_{\vec R} \right]_j  P(\vec R,t)  +T 
\sum\limits_i S^{-1}_{i,j} \partial_i  P(\vec R,t) \right\}  
+ \bar C(\vec R) P(\vec R) 
\end{equation}
where all the terms proportional to $P(\vec R)$ include  
$C(\vec R)$ and the ones that compensate the ones implied by the 
total divergence, namely $\bar C(\vec R)=C(\vec R) + \sum_j \partial_j \left[ 
S^{-1} f_{\vec R}\right]_j$. 
In the above equation $\bar C(\vec R)$ has not been computed explicitly  
as it should simply vanish because  it is determined by  
the standard property of the Fokker-Planck equation, namely that 
 the RHS should be 
a total divergence, so that once integrated over all volume, it guarantees that 
the normalization of the probability is conserved for any initial probability 
guess, as a simple consequence that the conditional probability satisfies 
$\int d[R^\prime]^N K(\vec R^\prime| \vec R) =1$ for any $\Delta$ and in particular in the limit $\Delta \to 0$.
Therefore  we finally obtain the following Focker-Planck equation with 
$\bar C(\vec R)=0$:
\begin{equation}
\partial_t P(x,t)= \sum\limits_j \partial_j 
\left \{ - \left[ S^{-1} f_{\vec x} \right]_j  P(x,t)  +T 
\sum\limits_i \left[ 
S^{-1}_{j,i} \partial_i  
P(x,t)\right]     \right\}
\end{equation}
that concludes the proof of this section.
  
\subsection{ Faster  iteration scheme}
 The previous Markov chain given in Eq.(\ref{smartMarkov}) solves the problem of computing the 
''cumbersome  term'' $\Gamma_j (\vec R)$ at the expense of computing the matrix $S(\vec R)$ two times for each iteration. 
In the following we describe 
another way to obtain the same Fokker-Planck equation, with an iterative 
scheme requiring only 
one evaluation of the matrix $S(\vec R)$. This  is important 
in our implementation of the Langevin dynamics, 
 whenever the 
evaluation of the matrix $S(\vec R)$ requires most computational effort 
as in the present application based on quantum Monte Carlo.
To this purpose the   
more convenient  iteration scheme defines  the new coordinates 
$\vec R(t+\Delta)$ not only in terms of $\vec R(t)$ but also 
of the previous one $\vec R(t-\Delta)$. 
This remains formally a Markov chain 
in an extended space acting on a $2N-$ dimensional 
vector  $\vec {\bf R}_n = [ \vec R_n , \vec R_{n-1} ]$, so that all the 
results of Markov chains used in the previous section can be used.
We propose therefore  the following iteration scheme: 
\begin{eqnarray} \label{finaldyn}
\vec R (t+\Delta) &=& \vec R(t) + \Delta S^{-1}\left(\vec R(t)\right) \vec f_{\vec R(t)} +\sqrt{2 T \Delta} \vec z (t) - {1 \over 2} S^{-1}\left(\vec R(t)\right) \left[ S\left(\vec R(t-\Delta)\right)- S\left(\vec R(t)\right) \right]  \left(\vec R(t-\Delta) -\vec R(t)\right) 
 \nonumber \\
\langle z_i (t) z_j (t) \rangle &=& S^{-1}_{i,j} (\vec R(t)) 
\end{eqnarray}
The important thing is to show that, for $\Delta\to0$ and 
 at the leading order, the previous 
iteration scheme is equivalent to the Markov chain in Eq.(\ref{noisez}).
This is evident by considering that, as in the previous case:
\begin{equation}
(\vec R(t-\Delta) -\vec R(t))=-\sqrt{2T \Delta} \vec z(t-\Delta)+O(\Delta) 
\end{equation}  
Therefore the ''cumbersome term'' in Eq.(\ref{finaldyn}) comes naturally from simple Taylor expansion:
\begin{equation} \label{cumber}
\left\{ 
\left[ S(\vec R(t-\Delta))- S(\vec R) \right]  (\vec R(t-\Delta) -\vec R(t))  \right\}_k
= 2T \Delta \sum_{i,l} \left(\partial_i S_{k,l}(\vec R(t))\right) z_i(t-\Delta)  z_l(t-\Delta)\simeq
2T \Delta \sum_{i,l}  \left(\partial_i S_{k,l}(\vec R) \right) z_i(t)  z_l(t)
\end{equation}
Thus the iteration scheme in Eq.(\ref{finaldyn}) is equivalent to the
following Markov chain:
\begin{eqnarray} \label{standardnew}
\vec R (t+\Delta)_j &=&  \vec R(t)_j + \Delta \left(S^{-1}(\vec R) \vec f_{\vec R} \right)_j 
 -T \Delta \sum\limits_{i,k,l} S^{-1}_{j,k}(\vec R) \left(\partial_i S_{kl}(\vec R)\right) z_i(t) z_l(t) 
 +\sqrt{2 T \Delta} z_j(t) \nonumber \\
\langle z_i (t) z_j (t) \rangle &=& S^{-1}_{i,j} (\vec R(t)) 
\end{eqnarray}
Strictly speaking the 
rightmost  equality in Eq.(\ref{cumber}) is valid for the associated Fokker-Planck equation where
one can substitute the correlator $\langle y_i(t-\Delta) y_l(t-\Delta) \rangle$
with the one $\langle y_i(t) y_l(t) \rangle$ with an error $O(\Delta)$.
Thus Eq.(\ref{standardnew})  
 is in turn equivalent to the standard covariant iteration scheme defined in Eq.(\ref{standard}) \cite{mazzola}, that is indeed obtained by substituting:
\begin{equation}
 z_i(t) z_l(t) \to S^{-1}_{i,l}(\vec R)= S^{-1}_{l,i} (\vec R)
\end{equation} 
in Eq.(\ref{standardnew}), 
that does not change the Fokker-Planck equation,
because this iteration scheme in Eq.(\ref{standardnew}) 
 coincides with the one we have considered in the previous 
section up to order $O(\Delta)$  
and therefore should lead to the same Fokker-Planck equation for $\Delta \to 0$. 
\subsection{ Better integration scheme, the adopted one}
Suppose that the Hessian matrix $H$ is proportional or almost proposrtional to the chosen matrix $S$, so that $H=\alpha S$ is a good approximation. Then, 
since $\vec f(\vec R) = -H (\vec R-\vec R_{eq})$, $\vec R_{eq}$ being the equilibrium position,  one can assume that:
\begin{equation}
S^{-1} \vec f (\vec R) = -\alpha \vec R + \vec C
\end{equation}
where $\vec C$ is an almost constant vector, as it depends only on the equilibrium position and the anharmonic terms.
Indeed, let us consider, for the time being, that we are in the harmonic 
case so that both $S$  and $\vec C$ do not depend on $\vec R$,
and Eq.(1) in the text simplifies as follows:
\begin{eqnarray} \label{basic}
\dot { \vec R} &=& -\alpha \vec R + \vec C 
  + \vec \eta(t)  \label{above} \\
\langle \eta_i \eta_j \rangle&=&2T  \delta(t-t^\prime) S_{ij}^{-1} \label{corrnoise}
\end{eqnarray}
Since the above Eq.~(\ref{above}) is now linear can be integrated exactly
in the interval $(t,t+\Delta)$ 
for arbitrary time dependency of the noise vector 
$\vec \eta(t)$, yielding:
\begin{eqnarray}
 \vec R(t+\Delta) -\vec R(t)  &=& (\exp( - \alpha \Delta ) -1) \vec R(t) +\int\limits_{t}^{t+\Delta} \exp\left[\alpha (\tau-t-\Delta)\right] 
\left[\vec C+ \vec \eta(\tau) \right] d\tau \nonumber \\
  &=& \tilde \Delta (-\alpha \vec R + \vec C)  + \sqrt{2 T \bar \Delta } \vec z(t) = \tilde \Delta 
S^{-1} \vec f (\vec R) 
 +\sqrt{2 T \bar \Delta } \vec z(t) 
  \label{eqsss}  \\
\sqrt{ 2 T \bar \Delta} \vec z(t) &=& \int\limits_{t}^{t+\Delta} \exp\left[\alpha (\tau-t-\Delta)\right] \vec \eta(\tau) \nonumber \\
\tilde \Delta &=& {1-\exp(-\alpha \Delta) \over \alpha } \nonumber \\
\end{eqnarray}
Moreover, by using Eq.(\ref{corrnoise}) one can derive 
the value of $\bar \Delta$, 
because:
\begin{equation}
2 T \bar \Delta \langle z_i(t) z_j(t) \rangle = 2 T
\int\limits_t^{t+\Delta}\int\limits_t^{t+\Delta}  \exp[  (\tau -t-\Delta) \alpha ]
\exp[  (\tau^\prime -t-\Delta) \alpha ]
 S^{-1}_{ij} \delta( \tau -\tau^\prime) d\tau d\tau^\prime
= 2T S^{-1}_{ij} \int\limits_t^{t+\Delta} \exp[ 2 (\tau-t-\Delta) \alpha ] d \tau
\end{equation}
implying that:
\begin{eqnarray}
\langle z_i(t) z_j(t) \rangle &=& S_{i,j}^{-1}  \nonumber \\ 
\bar \Delta &=& {1-\exp(-2\alpha \Delta) \over 2 \alpha } \nonumber \\
\end{eqnarray}

In the general case when also anharmonic terms are present and both $S$ and $\vec C$ are explicitly dependent on 
$\vec R$ it is useful to adopt the following choice that will match with Eq.(\ref{eqsss}) in the harmonic case
and therefore will be free of  
 time step error if the potential energy surface 
is well approximated by an  harmonic potential 
in the neighborhood of  $\vec R$: 
\begin{eqnarray} \label{finaldyn_new}
\vec R (t+\Delta) &=& \vec R(t) + \tilde \Delta S^{-1}\left(\vec R(t)\right) \left[ \vec f_{\vec R(t)}- {1 \over 2 \bar \Delta} \left[ S\left(\vec R(t-\Delta)\right)- S\left(\vec R(t)\right) \right]  \left(\vec R(t-\Delta) -\vec R(t)\right) \right] +\sqrt{2 T \bar \Delta} \vec z (t) 
 \nonumber \\
\langle z_i (t) z_j (t) \rangle &=& S^{-1}_{i,j} (\vec R(t)) 
\end{eqnarray}
Notice that we have used the discrete time step 
$\bar \Delta$ in the finite difference expression
containing $ {\left[ S\left(\vec R(t-\Delta)\right)- S\left(\vec R(t)\right) \right] \over \bar \Delta}$, because in the limit of $\bar \Delta \to 0$ this will restore the 
so called  ''cumbersome term'' in Eq.(\ref{e:cumber}),
with the correct prefactor. 
In the practical implementation the constant $\alpha$ used in the hydrogen case, when using the covariance matrix 
for $S=cov(\vec f)$, was set to:
\begin{equation}
\alpha= 0.122 N_c
\end{equation}
where $N_c$ is the number of QMC sampling used for  
each step of MD, in order to evaluate stochastically the 
force components.
The reason to scale the constant $\alpha$ by the number 
of QMC sampling is because the Hessian is not dependent obviously on $N_c$ whereas the covariance matrix, as it determines the squared stochastic fluctuations of the forces, has 
to decrease as $1 \over N_c$, 
by the central limit theorem. 
 
\subsection{ Noise correction}
In this QMC approach, we use  $S=Cov(f)$, 
where $Cov(f)$ is the correlation matrix corresponding to the statistical 
fluctuations -i.e. the error bars- of the nuclear forces.
In this case it is also very simple 
to correct for the extra 
noise given by the QMC forces\cite{yeluo_2014}. This is achieved 
in a very simple way, just by changing the 
temperature $T$ used in the dynamics for the correct simulation at a given target 
temperature $T_{target}$, simply as follows:
\begin{equation}
2 T \Delta = 2 T_{target} \Delta -\Delta^2. 
\end{equation} 
that is possible for $\Delta < 2 T_{target}$ (notice that for this particular choice of $S$ the time of the dynamics has the unusual dimension of an energy).
 In all simulations presented in this work, this correction was always negligible, i.e. less than $5\%$, which should represent an upper bound of our error in the equilibrium temperature.
In this work we have not used this correction, because, 
for the time step used, this correction is negligible.

\subsection{A simple toy model}
We have tested this dynamics in a simple toy model, a rotating spring lying on a plane
The endpoint $(x,y)$ is subject to a radial harmonic potential of the form
\begin{equation} \label{pot}
   U(x,y) = \frac12 k ( \sqrt{x^2+y^2}- a)^2~,
 \end{equation}  
while being free to rotate around the origin. 
 The configuration's space visited during the dynamics is a circular ring, whose radius is $a$ and width given by the (radial) thermal fluctuation.
This model is a prototypical example in which a strong decoupling of time scale, vibrational and rotational, is present.

Let us define the following matrix $G_\lambda$:
\begin{equation}
G_\lambda = \frac 1 \lambda \left(
\begin{array}{cc}
 \frac{  x^2+y^2 \lambda}{x^2+y^2} & \frac{x y (1 -\lambda)}{x^2+y^2} \\
 \frac{x y (1 - \lambda )}{x^2+y^2} & \frac{x^2 \lambda +y^2 }{x^2+y^2}
\end{array}
\right)
\end{equation}
where $\lambda$ is control parameter, and $G$ reduces to the identity when $\lambda = 1$.
It can be shown, following geometrical considerations, that, if $G_\lambda^{-1}$ multiplies the forces, it effectively reduces by a factor $\lambda$ the radial  component of the associated displacement.
We therefore use in $S=G_\lambda$ in Eq.~\ref{standard}, with $\lambda <1$ and $\gamma=0$.


\begin{figure*}[t]
\begin{minipage}{.48\textwidth}
    \includegraphics[width=3.2in]{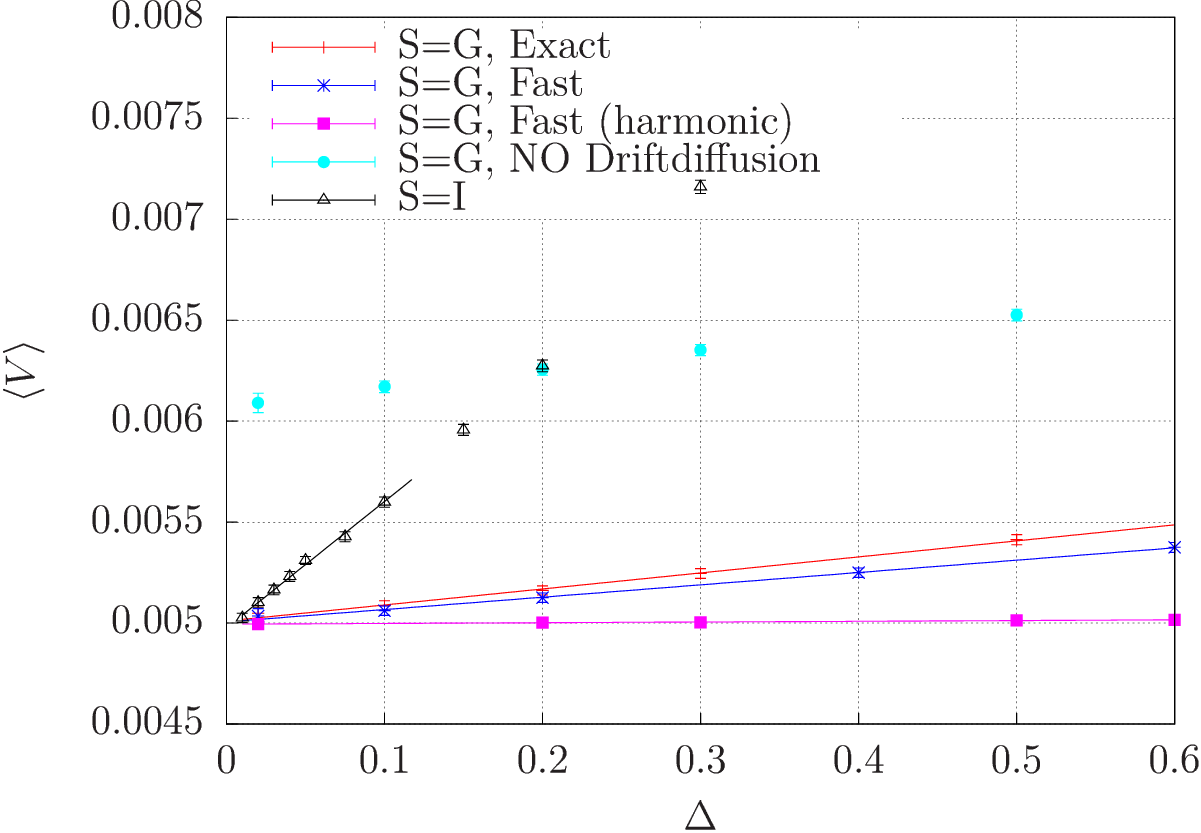}  
\end{minipage}
\begin{minipage}{.48\textwidth}
        \includegraphics[width=3.2in]{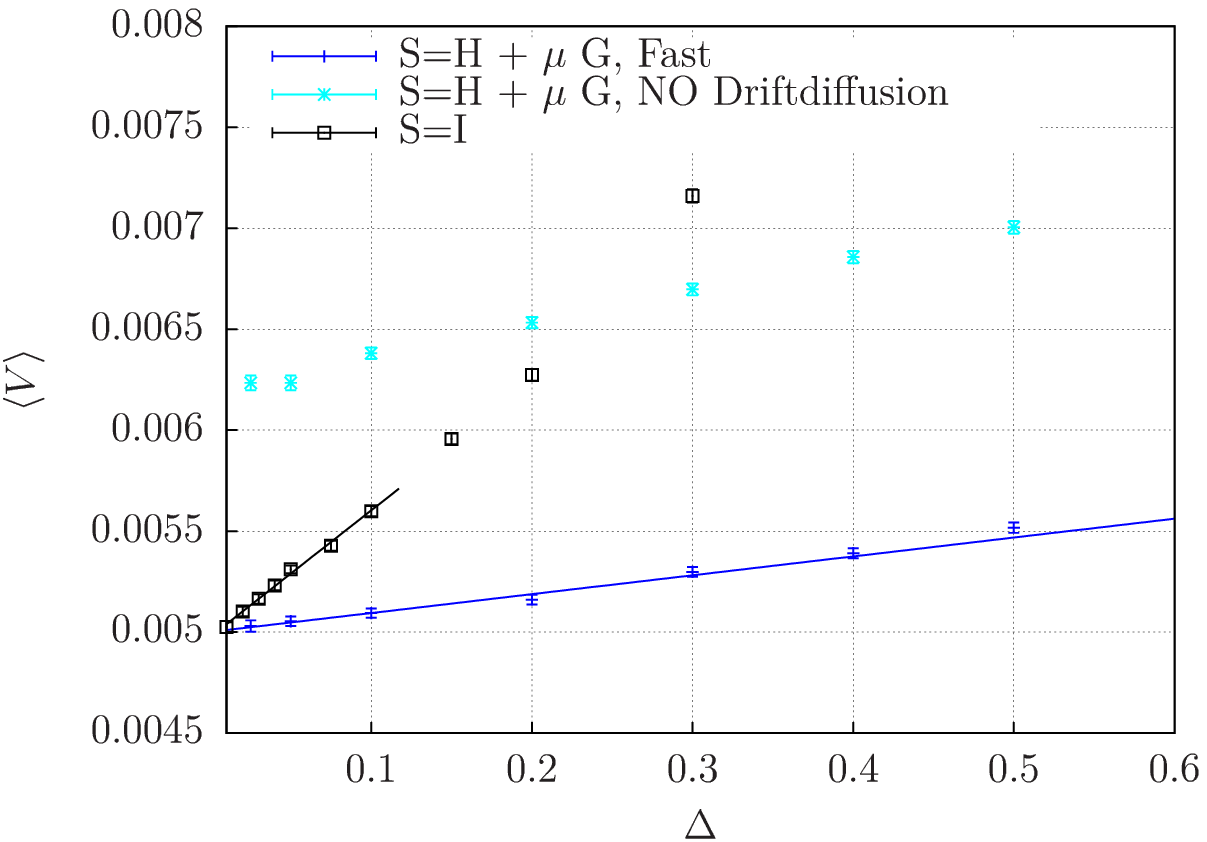}
\end{minipage}
 \caption{ Average potential energy as a function of the integration time-step $\Delta$. Black points refer to the standard first order Langevin dynamics, while colored points to the improved one, with non-trivial $S$. 
In the Left panel we use $S=G_\lambda$, while in the Right panel we use a combination of the Hessian matrix $H$, regularized with $\mu G_\lambda$ in order to have always a positive definite matrix $S=H +\mu G_\lambda$. 
 We use the following parameters: $\lambda=10, T=0.01, k=2, a=1.4$, and $\mu = 1/2$ in the Right panel.
We see that the new approach greatly alleviates the time-step error compared to the standard dynamics, while maintaining a very similar rotational diffusion coefficient at fixed $\Delta$ (not shown). 
Blue points refer to simulation obtained with Eq.~\ref{finaldyn}, while red ones to the exact integration of Eq.~\ref{standard}, computing explicitly the difficult term in Eq.~\ref{e:cumber}.
Magenta points correspond to Eq.~\ref{finaldyn_new} which is the integrator used for the hydrogen system. In this case we adopt $\alpha=0.2$. With this apprach we basically get rid of the time step error.
Solid lines represent linear fit of the respective data series.
In order to highlight the importance of the drift-diffusion term in ~Eq.~\ref{e:cumber}, we also perform simulations  using Eq.~\ref{norogna}. In this case the distribution sampled is not the correct one.
}
 \label{fig:toy}
\end{figure*}

From Fig.~\ref{fig:toy} we see that the preconditioned Langevin dynamics, with non-trivial $S$, results in a better time-step error, compared to the standard Langevin dynamics.
This demonstrate also in a simple toy model that a large computational gain can be achieved by this framework.
We use two different choices for the matrix $S$. In the first $S=G_\lambda$, while in the second $S=H +\mu G_\lambda$, where $H$ is the Hessian matrix.
In both cases, the correct equilibrium value is sampled in the $\Delta =0$ limit.

We also check the accuracy of the fast iteration scheme (Eq.~\ref{finaldyn}), against the exact one, in which we explicitly evaluate the drift-diffusion term in Eq.~\ref{e:cumber}.
The fast implementation extrapolates to the exact value and the time step error is comparable to the one produced by the exact dynamics.

Notice that, this term can be very important to sample the correct equilibrium distribution. 
Indeed, if we implement the simpler iteration rule
\begin{eqnarray} \label{norogna}
\vec R (t+\Delta) &=& \vec R(t) + \Delta S^{-1}\left(\vec R(t)\right) \vec f_{\vec R(t)} +\sqrt{2 T \Delta} \vec z (t) 
 \nonumber \\
\langle z_i (t) z_j (t) \rangle &=& S^{-1}_{i,j} (\vec R(t)) 
\end{eqnarray}
then the sampled distribution is simply wrong.

\clearpage
\section{Comprehensive results of the ab-initio simulations}
\subsection{Equations of state}
\begin{figure}[ht]
\begin{center}
\includegraphics[width=0.66\columnwidth ,angle=0]{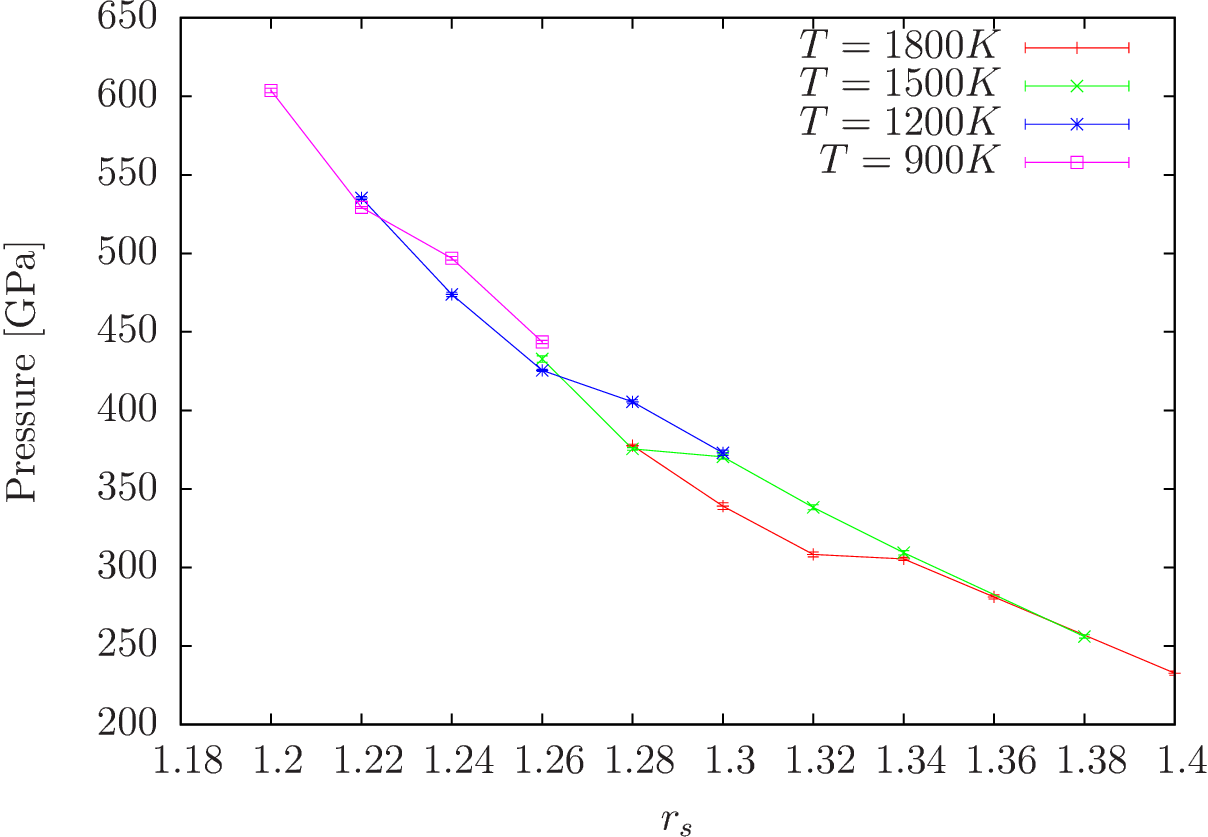}
\caption{ 
Equations of state $P$ vs $\rho$, for different temperatures: 900, 1200, 1500 and 1800 K 
The approximate discontinuity in the curves, as well as the $g(r)$'s change, allows us to identify the dissociation transition.
}
\label{fig:eqofstate}
\end{center}
\end{figure}
\clearpage
\subsection{Proof of equilibration of the MD}

Here we report all the outcomes of the simulations used to draw the hydrogen phase diagram.
In particular we demonstrate that the simulations are well equilibrated, and the results do not depend on the particular starting configuration (atomic or molecular).

A notable exception is represented by the few simulations performed at the proximity of the phase transition. Here, we can observe oscillations between the two phases within the same molecular dynamics run.
This further demonstrate the ergodicity of the simulations.

In the following we plot the pressure as a function of the simulation time (iterations).
The results refer always to a 64-atom system (see text).

\begin{figure}[h]
\begin{center}
\includegraphics[scale=1.1]{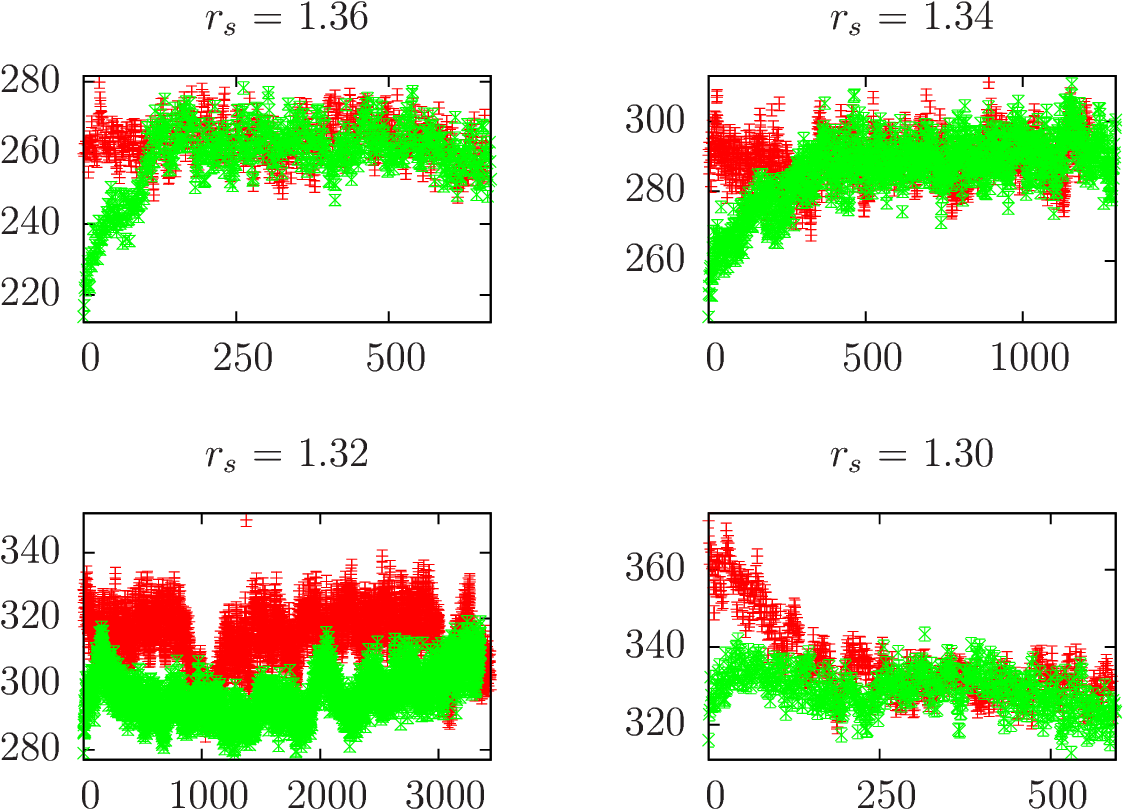}
\caption{{\bf T=1800 K .} Pressure (in GPa) as a function of the simulation time (in iteration) for different densities defined by Wigner -Seitz radius $r_s$ along the isotherm T=1800 K.  Red (green) points correspond to a MD which starting configuration is a molecular(atomic) liquid.}
\label{fig:p1800}
\end{center}
\end{figure}

Red (green) points correspond to a MD which starting configuration is a molecular(atomic) liquid. We see that, for fixed isotherm, at small densities (larger $r_s$), the MD which start from the full atomic configuration is the one far from equilibrium.
The opposite is true at larger densities (smaller $r_s$).
Near the phase transition (intermediate $r_s$) the two series meet halfway, although large oscillations are also present.
Notice for example the simulations at T=1800 K and $r_s=$1.32 in Fig.~\ref{fig:p1800}, which displays phase oscillations which periods of the order of thousand iterations.

%
%
%
%

\clearpage

\clearpage
\subsection{Radial pair distribution functions}

Here we report the radial pair distribution functions $g(r)$ for each isotherm, near the phase transition.
The $g(r)$ plot is useful to locate the phase boundary.
We plot in red (blue) simulations which produce a molecular (atomic) liquid. We compute the $g(r)$ discarding the initial equilibration steps.

\begin{figure*}[ht]
\begin{minipage}{.48\textwidth}
    \includegraphics[width=3.2in]{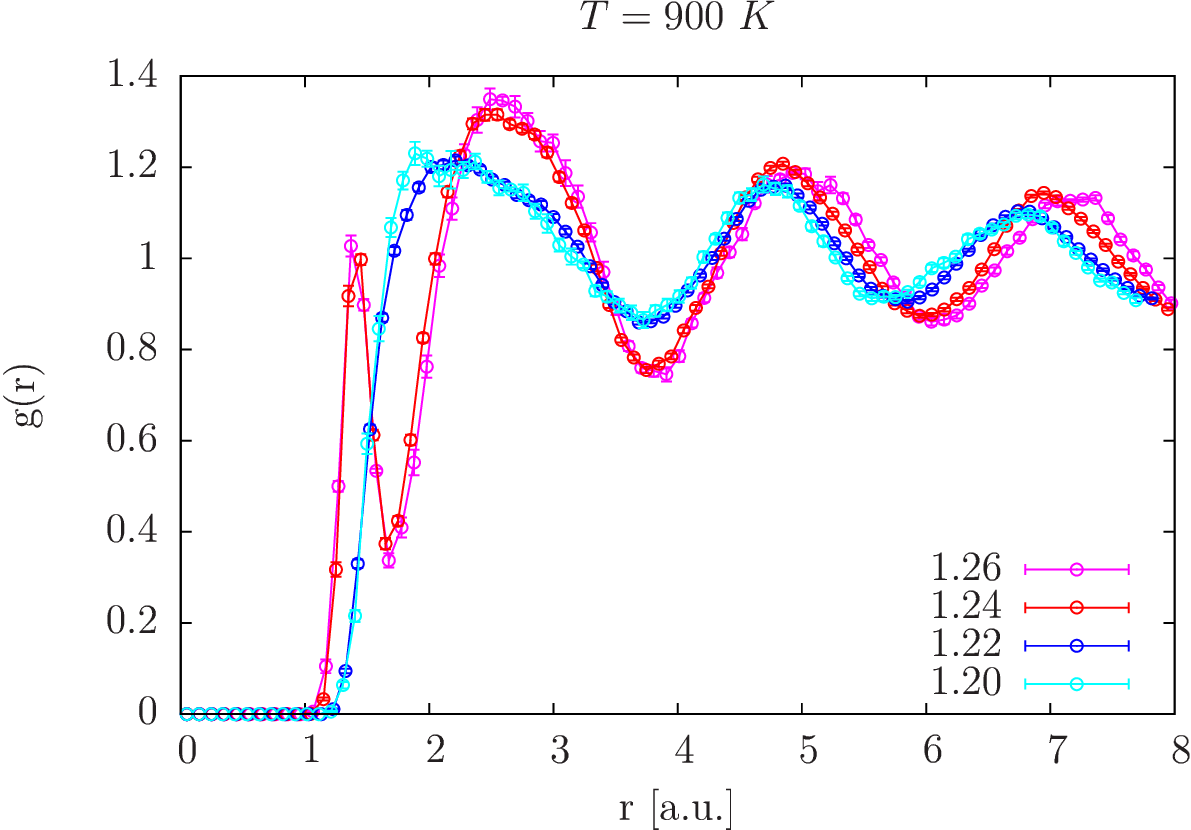}  
\end{minipage}
\begin{minipage}{.48\textwidth}
        \includegraphics[width=3.2in]{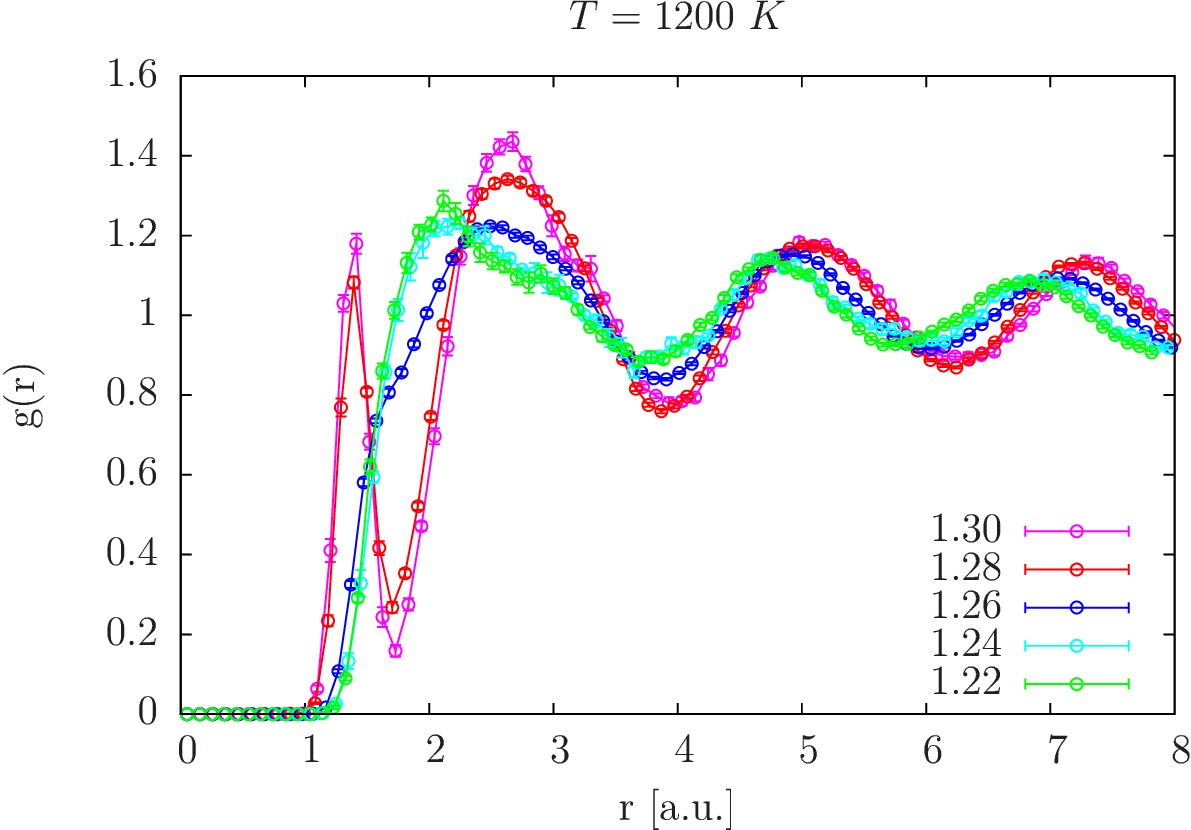}
\end{minipage}
\begin{minipage}{.48\textwidth}
        \includegraphics[width=3.2in]{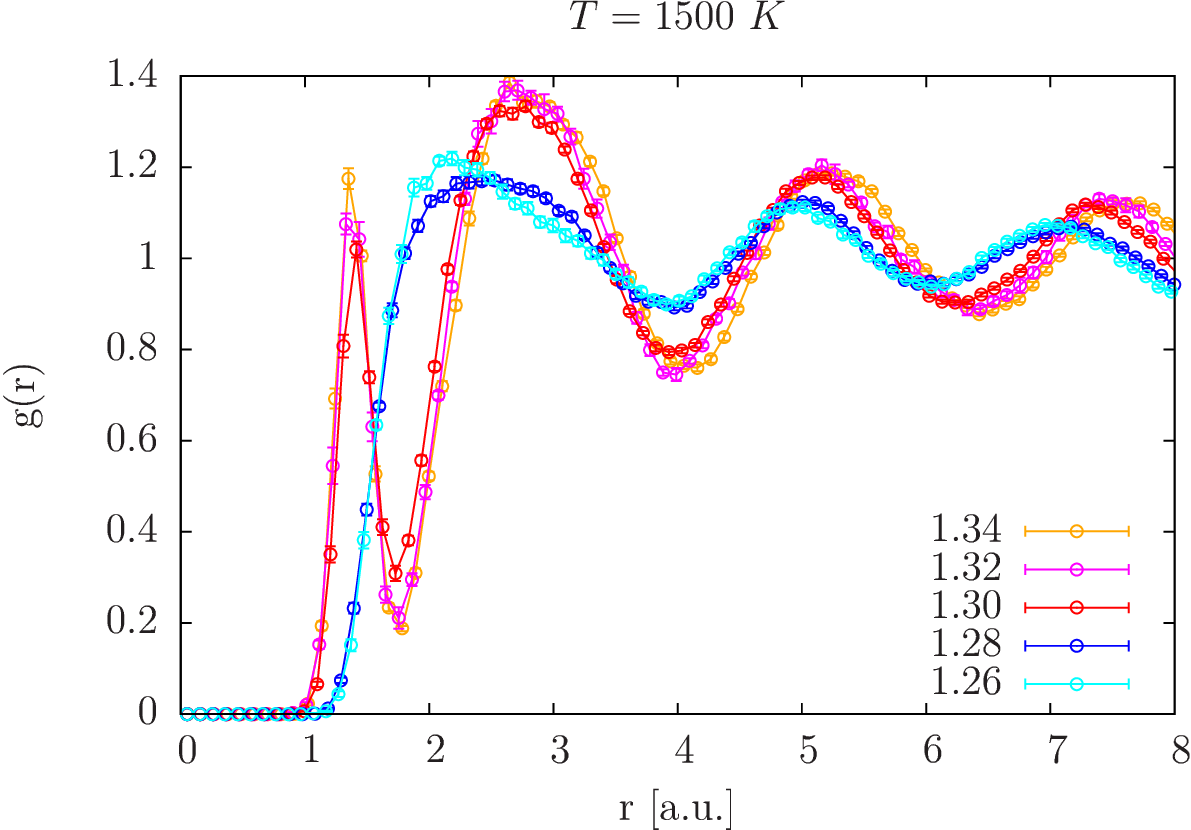}
\end{minipage}
\begin{minipage}{.48\textwidth}
        \includegraphics[width=3.2in]{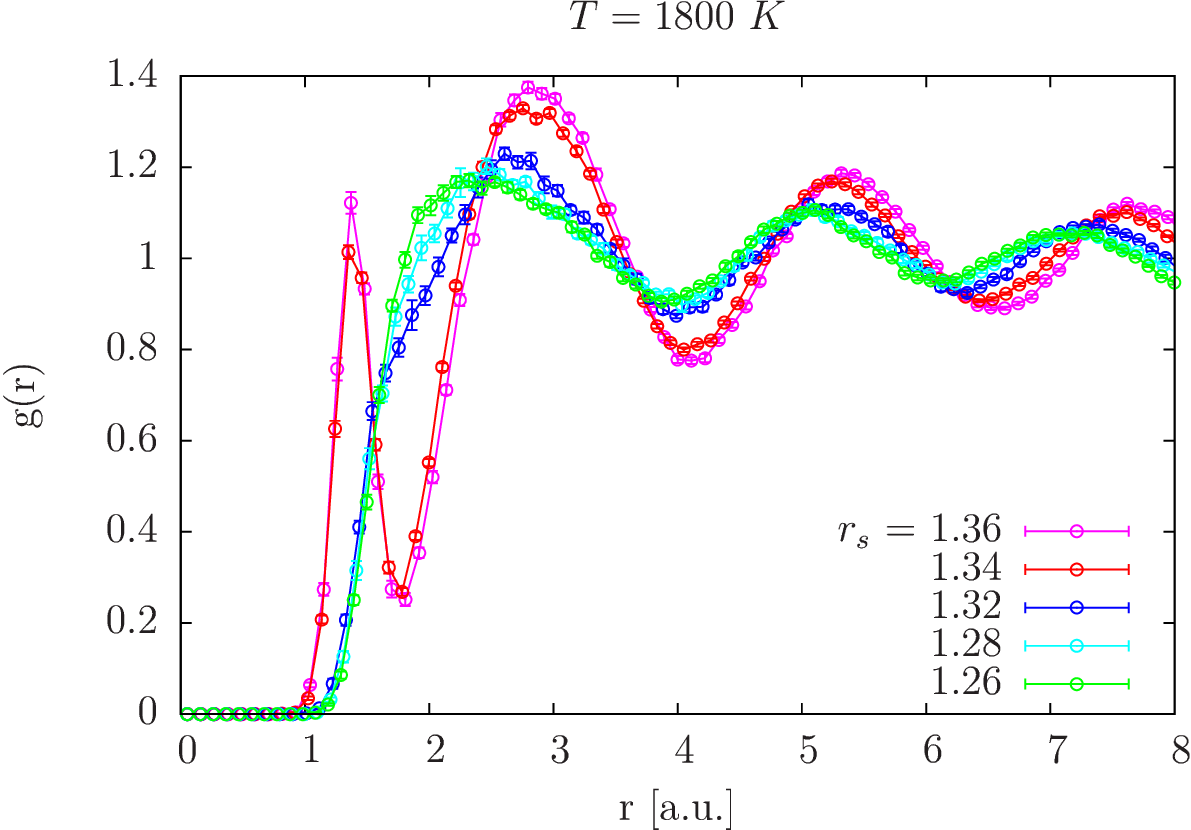}
\end{minipage}
 \caption{ Radial pair distribution function for different temperatures $T's$ and several densities ($r_s$) near the phase transition.
}
 \label{fig:grr}
\end{figure*}
%
\clearpage

\section{Finite size effects}

\begin{figure}[ht]
\begin{center}
\includegraphics[scale=1.1]{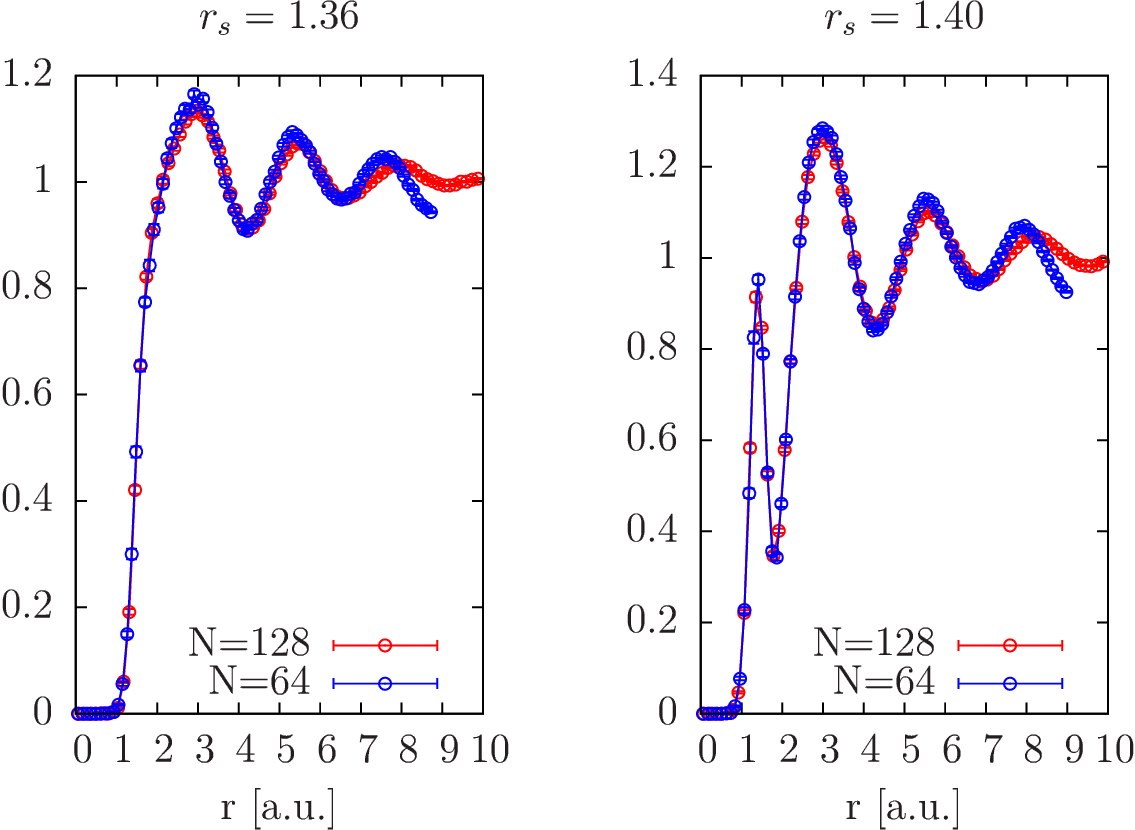}
\caption{{\bf Finite size errors.} Radial pair distribution functions at 1800 K near the phase transition ($r_s$=1.36 and 1.40) computed using N=128 (red) and 64 (blue) supercells. The finite size error is negligible in both cases.}
\label{fig:size}
\end{center}
\end{figure}

Finite size errors are negligible due to the twist boundary conditions.
Indeed, in Fig.~\ref{fig:size} we see that simulations using 64 and 128 atoms give essentially the same radial pair distribution functions.
The estimated pressures in all cases is very similar, the 128 atoms simulations giving a pressure larger than $\sim$ 10 GPa, compared to the 64 atoms ones.

\section{Systematic setup}

Each simulation in this work was done by starting from two randomly 
generated configurations of atomic and molecular character at $r_s^*=1.22$ ($r_s^*=1.27$) and $r_s^*=1.33$ ($r_s^*=1.31$) for the $64$ ($128$) proton case, respectively (see the corresponding files in this supplementary information). For all the other $r_s$ values these configurations were simply 
scaled by a factor $s={ r_s \over r_s^*}$.   
The minimal basis used for the determinant consists of one contracted 
orbital per atom. This orbital is a linear combination  of two Gaussians $ \phi_1(r)=  \exp(-Z_1 r^2) $ and $ \phi_2(r) =r^2 \exp( -Z_2 r^2) $ 
with:
\begin{eqnarray}
Z_1&=&  0.6066825   \\
Z_2&=&  0.3382168
\end{eqnarray} 
The coefficient $\eta$ of the linear combination $ \phi_1(r) +\eta \phi_2(r) $ 
is optimized on the fly during the run with all the other (up to several thousands)  parameters. 
We have also used the Gaussian exponents of a standard $ccp-VDZ$ basis\cite{basis} :
\begin{eqnarray}
 Z_{1s} &=& 1.962 \nonumber  \\
 Z_{2s} & = & 0.4446 \nonumber \\
 Z_{3s} &=& 0.122  \nonumber \\
 Z_{1p} &=& 0.727. \nonumber
\end{eqnarray}
In the standard basis, we have removed the Gaussian 
corresponding to  the largest exponent because in our approach the 
electron-nucleus cusp condition is properly taken into account by the presence 
of a one-body Jastrow factor\cite{zen}, that is also very useful  
for computing efficiently  the matrix elements of the  DFT calculation on  a finite mesh\cite{cava}.   
Notice that in this case we allow also a $p$ orbital of the standard Gaussian type (radial part $\phi_{1p}(r)=r \exp(-Z_{1p} r^2)$).
In order to minimize the number of parameters  
 this basis was contracted using 2 GEO atomic hybrid orbitals 
per atom\cite{geo}, instead of the standard, but much less efficient, atomic contraction for the 1s orbital.
This  is sufficient for a target accuracy in the DFT calculation of less than 0.5mH/atom, yielding a reduction of the number of parameters by a factor $\simeq 9$, as compared with the full uncontracted basis in the geminal expansion 
(see later). 
The coefficients of the contraction ($9$ independent parameters per atom) 
are optimized during the simulation. 
This basis set is used with generic twisted boundary conditions:
\begin{equation}
\phi(\vec r + \vec n L) = \exp( i \vec \theta \cdot \vec n) \phi(\vec r)
\end{equation}
where $\vec n$ is an integer vector and $\vec \theta$ is the corresponding 
 vector twist, whereas $L$ is the side of the cubic box.
The above Gaussian orbitals are modified in a way to satisfy the aforementioned  boundary conditions with standard methods\cite{crystal}.
The largest number of variational  parameters involved in the calculation 
comes from the geminal expansion $G(\vec r, \vec r^\prime)$ 
 in a localized basis :
\begin{equation}
G(\vec r_\uparrow, \vec r^\prime_\downarrow)  = \sum\limits_{i,j} \lambda_{ij} \phi_i (\vec r_\uparrow) 
\phi_j (\vec r^\prime_\downarrow) 
\end{equation}
All the $\lambda_{i,j}$ matrix elements can be considered free variational 
complex 
parameters to be optimized, corresponding to $p=2 L_A^2$ real parameters 
 where $L_A$ is the 
total single particle basis ($L_A=64$ for the smallest basis set 
with $64$H, and $L_A=128$  with the $2Z$ basis or $128$H).
$p=8192$ or $p=32768$ is already a too large number of parameters to be optimized efficiently within our statistical method. 
In the following we describe our strategy to reduce substantially (by about an order of magnitude) this number $p$. 
\begin{figure}[ht]
\begin{center}
\includegraphics[scale=0.6]{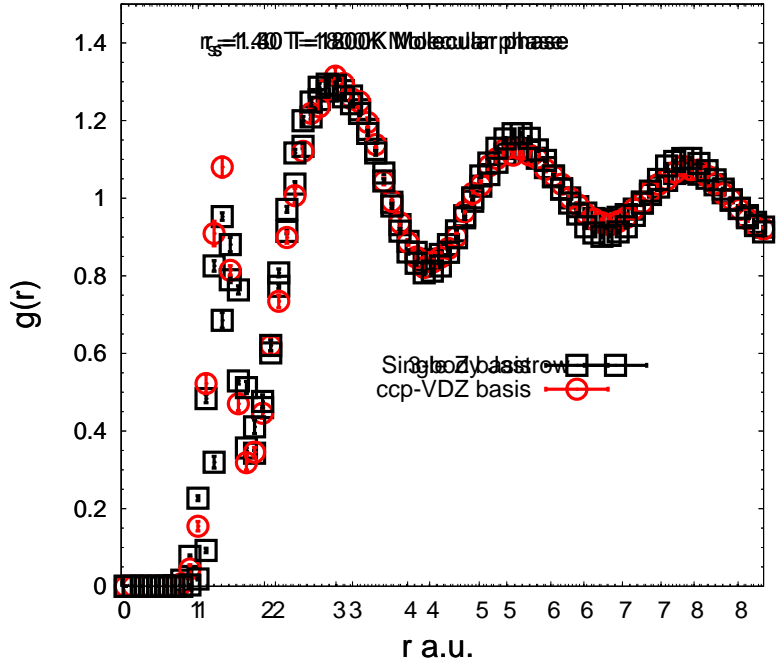}
\caption{{\bf Basis set dependence.} Radial pair distribution functions at 1800 K near the phase transition ($r_s$=1.40) computed using the two possible basis set for the determinantal part of the WF. The Jastrow is 3 body.}
\label{fig:basis}
\end{center}
\end{figure}
The Slater determinant can 
be described by a rank-deficient matrix 
 $\lambda_{i,j}$ 
with only $N/2$ non zero eigenvalues, where $N$ is the total number of 
electrons. 
The matrix $\lambda$ can be assumed hermitian provided 
 we adopt opposite twist vectors 
for opposite spin orbitals. This condition  allows the reduction of 
the number of parameters by a factor two within our projection method 
for the optimization of the determinant\cite{marchi}.
A further dramatic reduction of the number of parameters can be obtained 
in this approach by considering only variational parameters of the 
matrix $\lambda_{i,j}$ connecting localized orbitals at a distance less or 
equal than $3$a.u., and then projecting on the space of $N/2-$rank-deficient 
matrices, as described in details in Ref.\onlinecite{marchi}. 
We have checked that this approximation leads to negligible errors in the 
total energy, even for metallic Hydrogen.

The bosonic Jastrow term, $J = e^U$,
represents a compact way to  take into account explicitly the
electronic correlations since it depends directly on distances between electrons. There are many different choices for this factor; in this work we have used 
a Jastrow factor that accounts up to the 4-body interaction. The 1-body term is used to satisfy the nuclear-electron cusp condition. Hence the total Jastrow reads
\begin{equation}
J = J_1~J_2~J_3~J_4
\end{equation}

In  all cases studied the Jastrow basis is made of the same 
types of Gaussians orbitals $\phi_1(r)$ and $\phi_2(r)$ but without contraction 
for the three-body part and periodized with the trivial 
substitution\cite{spanu}:
\begin{equation}
 r_\mu \to {L\over \pi}  \sin ( { \pi \over L} r_\mu ).  
\end{equation}
When the electron-electron-ion-ion four-body part is used we contract this basis to one single orbital per atom only for this four-body term, whereas the full basis is used for the three body electro-electron-atom term, acting on two electrons around the same atom center.
The exponent used is taken the same for both orbitals $\phi_1(r)$ and $\phi_2(r)$, namely for the Jastrow: 
\begin{equation}
Z_1=Z_2=0.7933844
\end{equation}

In the following we provide the explicit functional forms of the Jastrow terms. Details can be found in Ref.~\onlinecite{marchi,zen} and references therein.
The term $U_1$ is a one electron interaction term which improves
the electron-nucleus correlation and satisfies the nuclear cusp
conditions. The exact functional form is given by
\begin{equation}
\label{e:j1}
U_1= -\sum_a^M [ (2Z_a)^{3/4} \sum_i^N u_1(\sqrt[4]{2 Z_a} ~r_{ia})] + \sum_a^M \sum_{\nu_a}^{L_a^J}\sum_i^N f^a_{\nu_a}\phi_{\nu_a}^a(r_{ia})
\end{equation}
where the vector $r_{ia} = r_i - { R}_a$ is the difference between the
position of the nucleus a and the electron i,  $Z_a$ is the electronic charge of the nucleus $a$,
 $L_a^J$ is the number of atomic orbitals $\phi_{\nu_a}^a$ that are used to describe the atom $a$,
$f^a_{\nu_a}$ are variational parameters and the function $u_1(x) = (1 - e^{-b_1 x})/2b_1$ depends parametrically on the value of $b_1$.

The $U_2(r)$ factor is an homogeneous two body interaction term. 
It depends only on the relative distance $r_{ij}$ between pairs of electrons. The specific functional form reads
\begin{equation}
 U_2(r) = \sum_{i<j}^N~u_2(r_{ij})
\end{equation}
where $u_2(x) = -{ x \over 2 ( 1 + b_2 x) }$
 and $b_2$ is a variational parameter.
Finally the 3 body term is an inhomogeneous two electron interaction that depends also on the relative position of the electrons and the nucleus, i.e. it's an \emph{e-e-n} interaction.
Its functional form is
\begin{equation}
\label{e:j33}
 U_3(r,{ R}) =  \sum_{i<j}^N [ \sum_a^M \sum_{\mu_a,\nu_a}^{L_a^J} f^a_{\nu_a,\mu_a}\phi_{\nu_a}^a(r_{ia}) \phi_{\mu_a}^a(r_{ja})     ]
\end{equation}
where $\phi_{\mu_a}^a$ are the uncontracted atomic orbitals centered on atom $a$.
Notice that this is an \emph{on site} interaction  which is included as a particular case, namely $a=b$ in the following equation,  of the more general 3-4-body $\emph{e-e-n-n}$ interaction
\begin{equation}
 U_{3+4}(r,{ R})=\sum_{i<j}^N [ \sum_a^M  \sum_b^M \sum_{\nu_a}^{L_a^J} \sum_{\nu_b}^{L_b^J} f^{a,b}_{\nu_a,\nu_b}\phi_{\nu_a}^a(r_{ia}) \phi_{\nu_b}^b(r_{jb})].
\end{equation}
In the genuine $4-$body term $U_4$ (with $a \ne b$) 
 we have used a 1s basis per atom,  after contraction 
of the 2s gaussian basis mentioned in the previous section. 
This allows the reduction of  the number of variational parameter in 
an  optimal way without significative loss of accuracy with respect to the full 
uncontracted 2s basis.

\subsection{ Hydrogen molecule }
\begin{figure}[ht]
\begin{center}
\includegraphics[width=0.66\columnwidth ,angle=0]{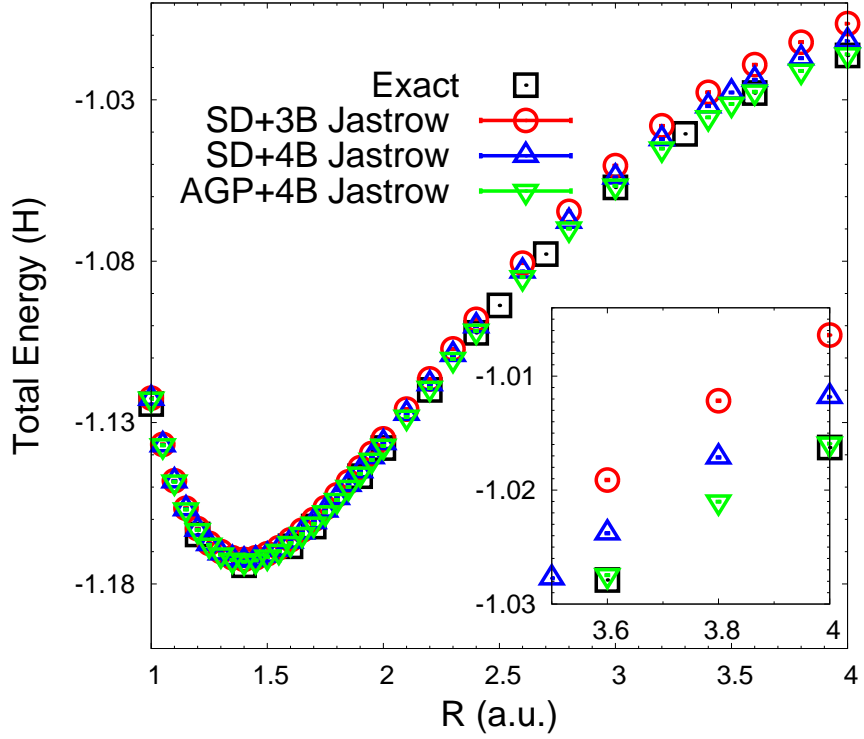}
\caption{ 
Energy as a function of the Hydrogen-Hydrogen distance in the Hydrogen molecule for three different wavefunctions as described in the text. The role of the 4-body (4B) 
Jastrow is particularly important at large distance for the single Slater determinant ansatz (SD). In this case also the exponents of the basis have been optimized.
}
\label{fig:dimer}
\end{center}
\end{figure}

In this section we show the dispersion curve of the Hydrogen molecule 
with the Jastrow single Slater determinant 
ansatzes used in this work (with and without the 
4-body Jastrow factor discussed in the last section).
 We also report in Fig.(\ref{fig:dimer}) 
the result of a Jastrow (4-body) single determinant 
Antisymmetrized Geminal Product (AGP)  wavefunction\cite{marchi}, in the same 
basis, showing that, even with a minimal basis 
without diffusive orbitals (p,d) both in the Jastrow and in the determinantal part, one can get an 
essentially exact description of the weak 
dispersive interaction in this molecule.
  
\end{document}